\documentclass[10pt,onecolumn,amsmath,amssymb]{article}
\usepackage{amsfonts}
\usepackage{amsmath}
\usepackage{amssymb}
\usepackage{graphicx}
\usepackage{bm}
\usepackage[super]{cite}
\usepackage{times}

\usepackage[font={small}]{caption}
\usepackage[title,titletoc]{appendix}

\newcommand{\onlinecite}[1]{\citen{#1}}
\newcommand{\refcite}[1]{\onlinecite{#1}}

\makeatletter
\renewcommand\section{\@startsection{section}{1}{\z@}%
{-3.5ex \@plus -1ex \@minus -.2ex}%
{2.3ex \@plus.2ex}%
{\normalfont\bfseries}}
\renewcommand\subsection{\@startsection{subsection}{1}{\z@}%
{-3.5ex \@plus -1ex \@minus -.2ex}%
{2.3ex \@plus.2ex}%
{\normalfont\bfseries}}
\makeatother

\date{}

\newcommand{\av}[1]{\langle{#1}\rangle}
\renewcommand{\v}[1]{|{#1}\rangle}
\newcommand{\iv}[1]{\langle{#1}|}

\renewcommand{\O}{{\cal O}}
\newcommand{\lr}{{\leftrightarrow}}

\begin{document}

\title{\large\bf
Quantum gates via continuous time quantum walks
\\\large\bf
in multiqubit systems with non-local auxiliary states}

\author{\normalsize
Dmitry Solenov
\\\normalsize\it
Department of Physics, St. Louis University, St. Louis, Missouri  63103, USA
\\\normalsize\it
solenov@slu.edu
}

\maketitle\thispagestyle{empty}

\begin{abstract}
Non-local higher-energy auxiliary states have been successfully used to entangle pairs of qubits in different quantum computing systems. Typically a longer-span non-local state or sequential application of few-qubit entangling gates are needed to produce a non-trivial multiqubit gate. In many cases a single non-local state that span over the entire system is difficult to use due to spectral crowding or impossible to have. At the same time, many multiqubit systems can naturally develop a network of multiple non-local higher-energy states that span over few qubits each. We show that continuous time quantum walks can be used to address this problem by involving multiple such states to perform local and entangling operations concurrently on many qubits. This introduces an alternative approach to multiqubit gate compression based on available physical resources. We formulate general requirements for such walks and discuss configurations of non-local auxiliary states that can emerge in quantum computing architectures based on self-assembled quantum dots, defects in diamond, and superconducting qubits, as examples. Specifically, we discuss a scalable multiqubit quantum register constructed as a single chain with nearest-neighbor interactions. We illustrate how quantum walks can be configured to perform single-, two- and three-qubit gates, including Hadamard, Control-NOT, and Toffoli gates. Continuous time quantum walks on graphs involved in these gates are investigated.

\vspace*{0.1truein}


\vspace*{0.1truein}

\noindent{\it Published}: 
Quantum Information and Computation {\bf 17}, 415 (2017)
\end{abstract}

\tableofcontents
\newpage

\section{Introduction}\label{sec:IN}

Quantum computing rely on two-state quantum systems (qubits) to store information and quantum gates to process it.\cite{DiVincenzo,Chakrabarti,nielsenchuang} Although other formulations exist, e.g., optical,\cite{Carolan} or measurement-based quantum computing,\cite{MeasureQI} this formulation has been one of the most commonly used due to its close analogy with classical binary information processing, among other reasons. One of the important elements of this analogy is design of quantum gates,\cite{nielsenchuang} which, in many cases, can be understood in terms of classical gate procedures applied to binary-labeled basis states. This means that quantum system must be driven by a classical external field to perform rotations of basis\cite{Louisell}---the amplitudes are driven or adiabatically carried through certain trajectories that start and end at some qubit basis states. In the case of entanglement-manipulating gates, such trajectories must also involve states that are formed due to physical interactions between qubits.\cite{nielsenchuang} These intermediate states, however, do not have to belong to qubits' computational basis.\cite{solenov1}

While qubits are binary quantum objects, physical systems that are used to represent them have more accessible quantum states. \cite{Schrodinger,Devoret,Wrachtrup,Koch,Awschalom,Carter} Additional higher-energy (auxiliary) states have been used in many architectures to manipulate qubits and develop entanglement.\cite{Koch,Steffen,Kim,Wrachtrup,Awschalom} Record coherence times and successful multiqubit manipulations recently achieved in systems of superconducting qubits that are nearly harmonic oscillators have brought this fact into focus once again.\cite{DiCarlo,Chow,Nigg,Rigetti,Paik} In these systems qubits are still encoded by the two lowest energy states. Yet, higher energy states are easily accessible and are not that distinct from the qubit states.\cite{Devoret,Nigg} It has been experimentally demonstrated that interaction via one of such higher energy states can be used to perform two-qubit entangling quantum gates in different quantum computing architectures, including those based on superconducting qubits\cite{Koch,Steffen} and self-assembled quantum dots.\cite{Kim} In all these cases the physics of performing entangling quantum gates involves driving the system through one non-local auxiliary state to accumulate a non-local phase for the wave function.

Recently we have demonstrated that a cavity-mediated interaction between multi-state quantum systems holding qubits generates a set of auxiliary states with certain structure of nonlocality that can be utilized to perform entangling quantum gates.\cite{solenov1,solenov2,solenov3} While this approach is applicable to more then two qubits, it suffers from spectral crowding and can become unusable for larger qubit systems.\cite{solenovNEW} In this paper we show that multiqubit systems interacting via multiple quantum fields (cavity modes) can overcome this difficulty. Under certain conditions local and non-local auxiliary states formed in these systems produce complex networks of states that do not suffer from spectral crowding and can be used to manipulate entanglement. We demonstrate that such networks can be effectively addressed if classical driving is replaced by temporarily-enabled continuous time quantum walks---a continuous time quantum evolution through a network of states with certain connectivity.\cite{Kempe,Kendon,Fedichkin1,Tamon,Fedichkin2,Zimboras} This approach gives multiqubit multi-state systems freedom to explore multiple quantum states involved in interactions, hence, potentially enabling more effective phase accumulation and faster quantum gates. We formulate general requirements on control and interactions between multi-state systems that are necessary to perform quantum gates via quantum walks in multiqubit registers. The procedure is illustrated with examples of one-, two-, and three-qubit gates.

The paper is organized as follows: Section~\ref{sec:QW} introduces continuous time quantum walk approach for multiqubit systems with multiple auxiliary states and interactions. In this section we formulate general requirements on control (driving) field---breaking of symmetry---needed to perform quantum gates on qubits via continuous time quantum walks. In Sec.~\ref{sec:Systems} we discuss realization of this symmetry breaking in scalable multiqubit architectures involving self-assembled quantum dots, defects in diamond, and superconducting transmon qubits. In Sec.~\ref{sec:QB} we show how this symmetry breaking can be utilized to perform quantum gates. We begin with the case of single-qubit gates in subsection \ref{sec:1QB}. In subsection \ref{sec:2QB} we formulate a class of quantum walk solutions representing Control Z gates (CZ, see Ref.~\refcite{nielsenchuang}). In subsections~\ref{sec:3QB}-\ref{subsec:sym-B}, systems of quantum walks performing diagonal Toffoli gates (Control Control Z, see Refs.~\refcite{nielsenchuang,Markov}) are obtained. Subsection~\ref{sec:time-comparison} is devoted to analysis of performance of walk-based gates. Detailed analytical investigation of continuous time quantum walks on all related graphs is given in the subsequent sections (Secs.~\ref{sec:LG}-\ref{sec:SQ}). Specifically, in Sec.~\ref{sec:LG} we investigate quantum walks on non-symmetric linear chain graphs with two to five nodes. In Sec.~\ref{sec:TG} we discuss quantum walks on single-level tree graphs. In Sec.~\ref{sec:SQ} we investigate quantum walks on symmetric and non-symmetric square graphs. Finally, concluding remarks are presented in Sec.~\ref{sec:conclude}.

\section{Quantum walks in qubit systems with states beyond boolean domain}\label{sec:QW}

Qubits are defined as binary (two-state) quantum systems.\cite{nielsenchuang} A distinction is often made between  logical qubits used in quantum algorithms\cite{Shor1,Zurek,Grover1,Shor2,Cleve,Grover2} and hardware-defined (physical) qubits that are parts of the physical system used for quantum computing. While this distinction is important because logical qubits can incorporate error correction procedures\cite{Aharonov,Steane,DiVincenzoShor,Kitaev,Rarity} based on operations involving multiple physical qubits, having a reliable set of entangling operations (gates) is crucial in both cases. Here we focus on physical qubits formed as parts of a larger quantum system,\cite{solenov3} each defined via Hamiltonian
\begin{eqnarray}\label{eq:QW:H_QB}
H^{(n)}_{QB} = E^{(n)}_0\v{0}\iv{0} + E^{(n)}_1\v{1}\iv{1}
\end{eqnarray}
Although not a matter of necessity,\cite{Pryadko} qubits are typically constructed such that they are well isolated from each other 
\begin{eqnarray}\label{eq:QW:H_QBs}
H_{QBs} = \bigotimes_{n=1}^{N} H^{(n)}_{QB}
\end{eqnarray}
to facilitate simpler error correction and algorithms development.\cite{nielsenchuang} We will focus on such case as it is relevant to many existing advanced qubit designs.\cite{Wrachtrup,Koch,Awschalom,Carter,Steffen} All of these physical systems naturally incorporate a set of well defined states beyond states $\v{0}$ and $\v{1}$ of each qubit. For many quantum computing designs these states are relied on for single-qubit rotations and initializations, and, in some cases, simple two-qubit manipulations.\cite{Wrachtrup,Awschalom,Carter,Steffen} When physical interaction between systems that encode qubits is present, these auxiliary states
\begin{eqnarray}\label{eq:QW:H_aux}
H_{aux} = \sum_{ij...\ne\!\!\!\mod 2} \varepsilon_{ij...}\v{ij...}\iv{ij...}
\end{eqnarray}
are not necessarily local to each qubit,\cite{solenov3}  i.e., $\v{ij...}\neq\v{i}\otimes\v{j}\otimes...$ for $ij...$ that have at least one non-binary digit (hence notation $ij...\ne\!\!\!\mod 2$). However, we will assume that states $\v{ij...}$ approach local states in the limit of no interaction between (physical) qubit systems. In that latter limit $\varepsilon_{ij...}\to E_i^{(1)} + E_j^{(2)} + ...$. This adiabatic connection will allow us to use the same labeling for interacting and non-interacting states to simplify further discussion. We will also assume that qubit states are not participating in any interaction (except with external control pulses) and remain local. 

The overall Hamiltonian of the system incorporating all relevant states is
\begin{eqnarray}\label{eq:QW:H}
H = H_{QBs} + H_{aux} + V(t)
\end{eqnarray}
where $V(t)$ represents external classical control\cite{Louisell} of the form
\begin{eqnarray}\label{eq:QW:V-gen}
V(t) = 2\Phi(t)\sum_{ij,\xi\xi'...}
\Omega_{i\xi\xi'...,j\xi\xi'...}
\v{i\xi\xi'...}\iv{j\xi\xi'...}
\cos(\omega_{i\xi\xi'...,j\xi\xi'...}t)
+ i.p. + h.c.
\end{eqnarray}
where $i.p.$ stands for index permutations, $\Phi(t)$ is a dimensionless pulse envelop function, and $\Omega$ are constant amplitudes of the corresponding harmonic of the control field. We assume that $\Phi(t)$ is slow relative to the carrier frequencies and has a single maximum, i.e., it represents the envelope function of a {\it single multicolor pulse}.

In order to eliminate local accumulation of phases due to, possibly distinct, qubit state energies $E^{(n)}_{i=0,1}$, we define qubits and focus on evolution in the rotating frame of reference (interaction representation\cite{nielsenchuang,vanKampen,Blum})
\begin{eqnarray}\label{eq:QW:H-int}
H_I(t) = e^{i (H_{QBs}+H_{aux}) t} V(t) e^{-i (H_{QBs}+H_{aux}) t}
\end{eqnarray}
In this case, $V(t)=0$ corresponds to trivial evolution (idling) of qubits, because qubit states do not participate in interaction. If $V(t)\neq 0$ for some period of time from $t_1$ to $t_2$, a non-trivial evolution (quantum gate) that involve one or more qubits and, possibly, interacting higher energy auxiliary states can occur. The corresponding evolution operator is
\begin{eqnarray}\label{eq:QW:Ug-gen}
U_g = P \left[ T\exp{-i\int_{t_1}^{t_2}dt H_I(t)}\right] P
\end{eqnarray}
where $T$ is time-ordering and $P$ is projection operator that projects onto qubit (boolean) domain  defined by Hamiltonian (\ref{eq:QW:H_QBs}). The projection signifies the fact that, ultimately, only qubit evolution is of interest: a leak from the qubit subspace can be a source of strong decoherence that is not addressable with standard error correction procedures. It is, therefore, crucial to ensure that $U_g$ is unitary 
\begin{eqnarray}\label{eq:QW:unitary}
U_g^\dag U_g = 1
\end{eqnarray}

In this paper we will focus on the case in which frequencies $\omega_{i\xi\xi'...,j\xi\xi'...}$ are in exact resonance with transitions in the system. In this case, dynamics leading to $U_g$ can be evaluated analytically: when rotating wave approximation is appropriate, the system can be mapped onto continuous time quantum walks on a graph with time-independent edges and nodes. To demonstrate this, note that within rotating wave approximation\cite{vanKampen,Blum}
\begin{eqnarray}\label{eq:QW:Hi-RW}
H_I(t)/\Phi(t) \to \Lambda = {\rm const}
\end{eqnarray}
and that the gate operator simplifies to
\begin{eqnarray}\label{eq:QW:Ug}
U_g \to P e^{-i \tau \Lambda} P
\end{eqnarray}
where
\begin{eqnarray}\label{eq:QW:tau}
\tau = \int_{t_1}^{t_2}dt \Phi(t)
\end{eqnarray}
is the effective time. 

Quantum computing is based on the principle that qubit amplitudes remain hidden (unknown) during the evolution (gates). As the result, quantum gates are designed to perform deterministic (classical) rotations of the basis, rather than change of amplitudes,
\begin{eqnarray}\label{eq:QW:evolve}
U_g \Psi = \sum_{ij...\in 0,1}\Psi_{ij...}\left[U_g\v{ij...}\right] 
\end{eqnarray}
Therefore, if we define $\psi(0) \equiv \v{ij...}$ and $\psi(\tau) \equiv U_g\v{ij...}$, quantum gate $U_g$ maps onto a set of continuous time quantum walks
\begin{eqnarray}\label{eq:QW:walk}
\psi(\tau) &=& e^{-i \tau \Lambda} \psi(0)
\end{eqnarray}
propagating in {\it effective time} $\tau$, where $\Lambda$ plays the role of a constant Hamiltonian or adjacency matrix (diagonal entries are zero in most cases) corresponding to a graph that defines each walk. This is in contrast with typical realizations of continuous time quantum walks discussed earlier,\cite{Fedichkin1,Tamon,Fedichkin2,Qiang} where propagation takes place in {\it real time}. Note that when rotating wave approximation is not appropriate, $\Lambda$ can still be defined, but it will become a function of time as well,\cite{Magnus,SolenovJMPB} in which case time-ordering must be honored.

To ensure conservation of probability within boolean (qubit) domain we must restrict ourselves only to a sub-set of graphs that satisfy
\begin{eqnarray}\label{eq:QW:QLP}
Q e^{-i \tau \Lambda} P = 0
\end{eqnarray}
where $P+Q=1$. In the trivial case when $P\Lambda P = \Lambda$ the walk never leaves the boolean domain (qubit subspace). Another important subgroup of graphs that satisfy condition (\ref{eq:QW:QLP}) is a set of graphs that 
enable ``{\it return}'' quantum walks---walks that return the population back to the initial state with probability 1 at some finite time $\tau$. 
In what follows we investigate graphs with $P\Lambda P \neq \Lambda$ that satisfy (\ref{eq:QW:QLP}). Particular emphasis is made on two types of {\it return} quantum walks: (i) walks that accumulate no phase when the population is returned to the original state (trivial return walks), and (ii) walks that accumulate a phase of $\pi$ when return to the initial state (non-trivial return walks). The simplest example of such walks is the evolution of a driven two-state quantum system.\cite{RosenZener}

The above description can be easily generalized to include multiple multi-color pulses, each performing its own kind and set of quantum walks. In this case Eq.~(\ref{eq:QW:V-gen}) is replaced with
\begin{eqnarray}\label{eq:QW:V-multi}
V(t) = V(t;\{\Phi^1,\Omega^1\}) + V(t;\{\Phi^2,\Omega^2\}) + ...
\end{eqnarray}
where a different set of Rabi frequencies, $\Omega^n$, can be chosen for each pulse $V(t;\{\Phi^n,\Omega^n\})$ to provide a more complex time-depended control. Examples of both single- and multi-pulse control will be given in later sections. Note that quantum walks corresponding to each pulse propagate in their own times
\begin{eqnarray}\label{eq:QW:tau-multi}
\tau^n = \int_{t^n_1}^{t^n_2}dt \Phi^n(t)
\end{eqnarray}
independently from each other. The gate operator is a product
\begin{eqnarray}\label{eq:QW:Ug-multi}
U_g \to P e^{-i \tau^1 \Lambda^1}\times e^{-i \tau^2 \Lambda^2} \times ... \,P
\end{eqnarray}
with projection, $P$, applied only twice---amplitudes in between the pulses do not have to reside in the qubit subspace. The total physical time span of the gate is
\begin{eqnarray}\label{eq:QW:t-total-multi}
\Delta t_{\rm total} = (t^1_2-t^1_1) + (t^2_2-t^2_1) + ...
\end{eqnarray}
Because quantum walk pulses can involve multiple non-equal Rabi frequencies that are effectively ``multiplied'' by the time duration $t^n_2-t^n_1$ of each pulse, some specific convention must be adopted to compare the duration of gates performed this way to the duration of gates or gate decompositions performed by single-frequency pulses.\cite{solenov1,solenov2,solenov3} For this purpose, it is natural to limit the maximum Rabi frequency of each pulse (pulse field amplitude) to some  value accessible to specific experimental setup (and the same for all pulses) and adjust $t^n_2-t^n_1$ to produce entries of the desired magnitude in each $\tau^n\Lambda^n$ matrix.

\section{Graphs and connectivity in scalable multiqubit systems}
\label{sec:Systems}

In the system introduced in Sec. \ref{sec:QW}, the adjacency matrix is a collection of complex Rabi frequencies originating from the single control pulse (\ref{eq:QW:V-gen})
\begin{eqnarray}\label{eq:QB:Lambda}
\Lambda = \sum_{\bf i,i'}\Omega_{\bf i,i'} \v{\bf i}\iv{\bf i'}
\quad\quad {\bf i}=ij...
\end{eqnarray}
The graph corresponding to this adjacency matrix is a set of vertices representing states $\v{ij...}$, connected via complex hopping amplitudes $\Omega_{\bf i,i'}$. Because these hopping amplitudes represent strengths of Fourier harmonics of external control field, they are adjustable parameters of the problem and can be chosen to perform the desired quantum walks and, ultimately, quantum gate. Not all these amplitudes, however, are independent.

When multi-state systems that hold qubits are well isolated from one another, a set of Rabi frequencies describing transitions in the system obeys strict symmetry relations. All graph node states $\v{ij...}=\v{i}\otimes\v{j}\otimes...$ are product states, and external control field can rotate each individual isolated multi-state system independently of the state of other such systems, i.e., 
\begin{eqnarray}\label{eq:QB:NI-Omega}
\nonumber
\Omega_{ijk..., i'jk...}
&=& \Omega_{ij'k'..., i'j'k'...}
\quad\quad
\forall jj'kk'...
\\
\Omega_{jik..., ji'k...}
&=& \Omega_{j'ik'..., j'i'k'...}
\quad\quad
\forall jj'kk'...
\\\nonumber
&...&
\end{eqnarray}
Note that all Rabi frequencies $\Omega$ in each row correspond to the same physical harmonic of the external control field resonantly driving transition $\v{i}\lr\v{i'}$ in the respective isolated qubit system.

This symmetry can be partially or completely lifted when there are physical interactions between parts of the system that encode different qubits, i.e., graph vertex states $\v{ij..}$ are no longer separable (factorisable) for some or any $i,j,...$. The degree of symmetry reduction depends on the strength of interactions as will be illustrated below for specific cases. Nevertheless, groups of indistinguishable $\Omega$-s may still exist if the spectrum has degenerate transitions corresponding to specific symmetries in the interacting system. In addition, degeneracy in graph edges (values of $\Omega$) can be artificially introduced, even if not present originally, by choosing appropriate amplitudes for the harmonics of the external control pulse.

The structure of the symmetry breaking that results in violation of relations (\ref{eq:QB:NI-Omega}) depends on the structure of interaction and also its strength. Particularly, in the case of small number of qubit systems coupled via a single quantum field, such as two qubits interacting via a single cavity, a specific dependence on the interaction strength was demonstrated\cite{solenov3}---an ``intermediate resonance regime''. It is realized when the cavity-induced interaction is weak to split degeneracy in some transitions as compared to pulse widths, but already sufficiently strong to lift it for other transitions in the system, hence partially lowering symmetry (\ref{eq:QB:NI-Omega}). As the result, local single-qubit gates and non-local entanglement manipulations can be performed by pulses without changing the strength of interactions or shifting qubits' energy levels dynamically.\cite{solenov1,solenov2,solenov3} Unfortunately, the intermediate resonance regime in the single-cavity system is not scalable beyond several qubits due to spectral crowding that hinders the distinguishability of states for realistic values of pulse bandwidth.\cite{solenovNEW}

In the following subsections we will show how symmetry breaking in relations (\ref{eq:QB:NI-Omega}) can occur for a scalable multiqubit register. We will focus on the approach that relies on multiple (orthogonal) cavity modes to carry interaction between qubit systems in the register, and will use principles of the intermediate resonance regime developed in our earlier work.\cite{solenov3} In order to provide examples, we will investigate three different qubit architectures that have demonstrated substantial experimental progress recently: self-assembled quantum dots, NV-center in diamond, and superconducting transmon qubits. The first system will be discussed in greater details introducing principles that will also be useful for the other two qubit architectures.

\subsection{Self-assembled quantum dots}
\label{sec:Systems:QDs}

\begin{figure}\begin{center}
\vspace*{1.1truein}
\includegraphics[width=0.99\columnwidth]{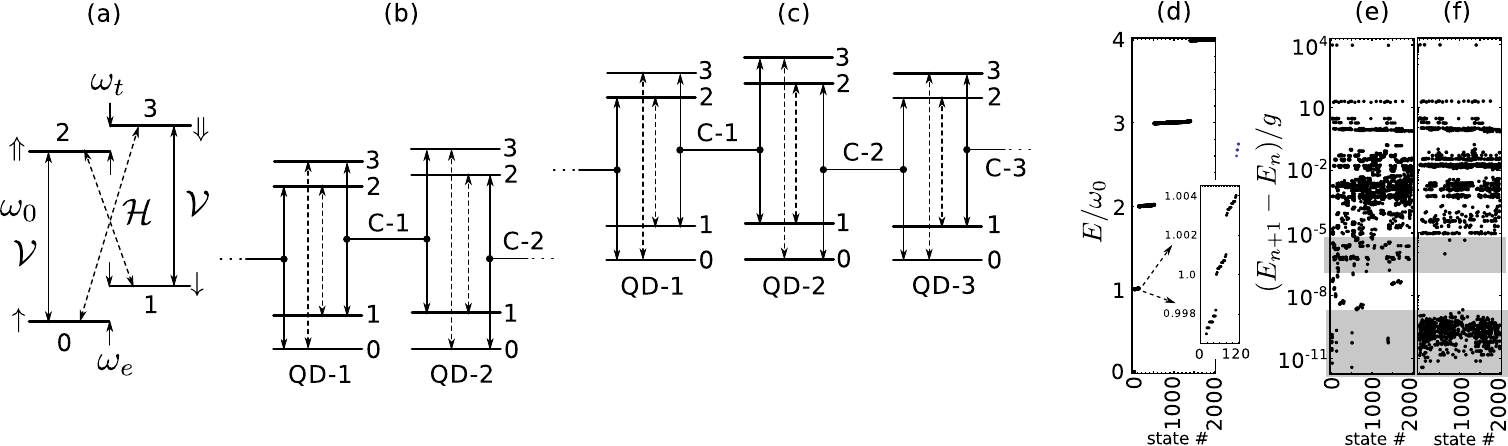}
\caption{\label{fig:QDs}
Level diagrams for a system of self-assembled InAs/GaAs quantum dots. (a) Relevant energy states of a negatively charged InAs/GaAs quantum dot. States $\v{0}\equiv\v{\uparrow}$ and $\v{1}\equiv\v{\downarrow}$ are electron spin states encoding a qubit. States $\v{2}\equiv\v{\Uparrow}$ and $\v{3}\equiv\v{\Downarrow}$ are optically accessible collective charged exciton (trion) states. Dashed, $\cal H$, and solid-line, $\cal V$, transitions are coupled to the two orthogonal light polarizations. (b) A segment of a scalable quantum register: two quantum dots (QD-$n$) connected via cavity modes (C-$n$). The cavity modes are coupled to the $\cal V$ polarization in this example. Transitions coupled to the same and the other polarization (dashed lines) can be activated in each dot with focused laser pulses. Excitonic transitions in quantum dots 1 and 2 must be spectrally distinct to avoid spectral crowding. (c) The segment shown in (b) connected to the next dot. Note that dots with the same (index) parity along the chain, i.e., QD-1, QD-3, QD-5, etc., can be identical. (d) Part of the spectrum of a chain of four dots coupled via three cavity modes, $\omega_0/g=10^4$. (e) Distances between neighboring energy levels in (d). (f) The same as (e), but with cavity modes artificially restricted to couple only to one transition in each dot to block propagation of excitations along the chain. The upper gray shading approximately outlines the range corresponding to translation-induced splittings. They disappear (except for few accidental degeneracies) on panel (f). The lower gray shading outlines limits of numerical diagonalization accuracy. 
}
\vspace*{1.1truein}
\end{center}\end{figure}

We begin with qubit systems based on self-assembled InAs/GaAs quantum dots.\cite{solenov1,Carter} In this systems qubits are encoded by an electron or hole spin ($\v{\uparrow}$ and $\v{\downarrow}$) corresponding to a state localized in the dot. Fast external control is achieved by optical driving of a negatively charged exciton, or a trion,---a collective excitation that carries an effective net angular momentum of 1/2 that can have both spin and orbital contributions (states $\v{\Uparrow}$ and $\v{\Downarrow}$). All relevant degrees of freedom of a single dot can be described by Hamiltonian
\begin{eqnarray}\label{eq:QDs:H}
H_{QD} = 
  E_\uparrow\v{\uparrow}\iv{\uparrow}
+ E_\downarrow\v{\downarrow}\iv{\downarrow}
+ E_\Uparrow\v{\Uparrow}\iv{\Uparrow}
+ E_\Downarrow\v{\Downarrow}\iv{\Downarrow}.
\end{eqnarray}
The $\uparrow/\downarrow$ and $\Uparrow/\Downarrow$ energies split in an external magnetic field as $E_\uparrow - E_\downarrow = \mu_e {\cal B} \equiv \omega_e$ and $E_\Uparrow - E_\Downarrow = \mu_t {\cal B} \equiv \omega_t$ with $\mu_t\neq\mu_e$, where g-factors have been included into the definitions of $\mu$-s. The corresponding level diagram is shown in Fig.~\ref{fig:QDs}(a). Because the primary control mechanism in the system is creation of charged exciton, we will refer to excited states as states with at least one exciton. The system can be controlled by coherent laser field coupled to excitonic transition\cite{Kim}
\begin{eqnarray}\label{eq:QDs:H}
V_{QD}(t) = 
  2\Phi(t)\left\{
  \Omega_{\cal V}\cos\omega_{\cal V}t
  \left(\v{\uparrow}\iv{\Uparrow} + \v{\downarrow}\iv{\Downarrow}\right)
+  
  \Omega_{\cal H}\cos\omega_{\cal H}t
  \left(\v{\uparrow}\iv{\Downarrow} + \v{\downarrow}\iv{\Uparrow}\right)  
  \right\}
+ h.c.,
\end{eqnarray}
where $\cal V$ and $\cal H$ denote two orthogonal polarizations of the laser pulse. We will assume that harmonics of the multicolor control pulse can be applied (focused) locally to each quantum dot. The real-valued dimensionless pulse envelop function $\Phi(t)$ is the same for all harmonics, as defined earlier in Eq.~(\ref{eq:QW:V-gen}). The relation between axises of polarization and the growth direction of the dots depend on several factors, such as light-heavy hole mixing, that are set, predominantly, at the time of manufacturing.\cite{Carter} Additional control can be achieved with microwave pulses coupled directly to spin states $\v{\uparrow}$ and $\v{\downarrow}$. The microwave operations however are typically slower than optical control. Note that while transitions $\v{0}\lr\v{2}$ and $\v{1}\lr\v{3}$ can be distinguished from $\v{0}\lr\v{3}$ and $\v{1}\lr\v{2}$ by polarization to which they couple, transitions $\v{0}\lr\v{2}$ and $\v{0}\lr\v{3}$ are distinguishable from $\v{1}\lr\v{3}$ and $\v{1}\lr\v{2}$, respectively, only spectrally.

Self-assembled quantum dots can be coupled to photonic crystal cavity modes,\cite{Carter} which act as coherent quantum medium to carry interaction between qubits. Because trion excitations can be distinguishable by polarization, two orthogonally-polarized cavity modes can, in principle, be set up to interact with ``spin conserving'' and ``spin-flip'' transitions independently\cite{Carter}
\begin{eqnarray}\label{eq:QDs:H}
H_{DC} = 
  \left(\v{\uparrow}\iv{\Uparrow} + \v{\downarrow}\iv{\Downarrow}\right)
  g_{\cal V}
  \left(a_{\cal V}^\dag+a_{\cal V}\right)
+  
  \left(\v{\uparrow}\iv{\Downarrow} + \v{\downarrow}\iv{\Uparrow}\right)  
  g_{\cal H}
  \left(a_{\cal H}^\dag+a_{\cal H}\right)
+ h.c.
\end{eqnarray}
Furthermore, multiple cavity modes with the same polarization but corresponding to different frequencies can couple to the same set of transitions at the same time, e.g., $g_{\cal V}a_{\cal V}\to g_a a + g_b b$. In what follows we will discuss the case $g_{\cal H}=0$ and will omit indexes in the coupling strength constant $g_{\cal V}\to g$ to simplify notation. We will also assume that $g$ is the same for all dots. Inhomogeneity in the coupling strengths at different quantum dots, unless significant, will not change the results qualitatively.

In general, the spectrum of $N$ identical or sufficiently similar quantum dots coupled in a chain, as shown in Figs.~\ref{fig:QDs}(b) and (c), has bands of states corresponding to propagation of excitations through the chain. When strength of coupling to cavity modes, $g$, is zero, these bands are degenerate states (zero band width), in which each state is local and distinguishable by the appropriate pulse harmonic of a pulse focused on specific dot. When $g$ is finite and $N\to\infty$, each individual state becomes spectrally indistinguishable because the states are no longer local---excitations propagate through the chain. In this case the band width is $\sim g$, and the number of states within each band is $\sim N$. As a result, states become spectrally indistinguishable for realistic pulses, which is referred to as ``spectral crowding.'' For example, a cavity photon from cavity C-$(2n-1)$ can be absorbed by transition $\v{0}\lr\v{2}$ in the right adjacent quantum dot, and then subsequently emitted as cavity C-$2n$ photon, and so on. This propagation of excitations can, in principle, be suppressed if odd and even cavity mode photons do not couple to the same transitions (modes themselves are orthogonal to each other). In such case the resulting spectrum would resemble that of quantum dot pairs, with each energy being highly degenerate if the pairs are identical. The degeneracy in this case is not a problem because each state is local to its pair of dots, and, hence, is addressable locally, i.e., is distinguishable.

Cavity modes connecting a chain of quantum dots will necessarily couple to each other via excitonic transitions unless they are sufficiently detuned in frequency. Detuning reduces coupling to individual transitions from $g$ to $\sim g/\delta\omega$ where $\delta\omega$ is the detuning energy, hence reducing the interaction. As we have demonstrate earlier,\cite{solenov1,solenov3} double-dot systems with finite detuning are still suitable for entanglement manipulations via excitonic states provided detunings between optical transitions in the dots, as well as the detunings between excitonic transitions and cavity photons, are within certain range as defined by intermediate resonance regime.\cite{solenov3}. In a system of $N$ dots a similar regime can, in principle, develop if all dots are detuned from one another.\cite{solenovNEW} This, however, is not practically achievable for large number of coupled quantum dots. Below we demonstrate that it is sufficient to detune only the nearest neighbor dots, while the next nearest neighbors can be similar. 

Consider a system in which optical transitions in the adjacent dots are detuned by $\sim\Delta \gg g$, while transitions in the dots with the same parity (of the index) are approximately equal to each other (with detuning $\lesssim g$). The latter condition can be relaxed and is chosen for clarity of presentation. In this case the chain is composed of identical (or similar) pairs of dots shown in Fig.~\ref{fig:QDs}(b). The even cavity modes are detuned to the blue by $\sim\Delta$ from the largest frequency $\v{1}\lr\v{3}$ transition, and the odd cavity modes are detuned to the red by $\sim\Delta$ from the smallest frequency $\v{0}\lr\v{2}$ transition, or vice versa [see Fig.~\ref{fig:QDs}(c)]. Which parity cavity mode is detuned to higher energies, as well as the specific value of detuning, will not be significant, but that choice and the order of magnitude for detunings, $\Delta$, must be the same for the entire register. This results in a $g/\Delta$ factor each time a cavity mode photon is absorbed or emitted, e.g.,
\begin{eqnarray}\label{eq:QDs:split}
\v{1010..,a_1^\dag}
\xrightarrow[]{\sim g/\Delta}
\v{1210..}
\xrightarrow[]{\sim g/\Delta}
\v{1010..,a_2^\dag}
\xrightarrow[]{\sim g/\Delta}
\v{1030..}
\xrightarrow[]{\sim g/\Delta}
\v{1010..,a_3^\dag}
\end{eqnarray}
Therefore the amplitude of translating the cavity excitation one step to the next equivalent cavity is $\sim (g/\Delta)^4$. The width of the energy bands resulting from such translations will be $\sim g(g/\Delta)^4$. The intermediate resonance regime for each pair of dots requires that transitions split by $\sim g^2/\Delta$ are distinguishable to the driving pulse, while transitions split by $\sim g(g/\Delta)^4$ are indistinguishable.\cite{solenov3} This means that shifts $\sim (g/\Delta)^4$ and possible resulting differences in excitonic transitions should appear effectively indistinguishable. Each such state remains effectively local to one of the quantum dot pairs as in the case discussed above when cavity modes did not couple to the same transitions. Therefore, despite of the translational symmetry along the chain (of base 2), spectral crowding will  not occur. At the same time, many non-local states that span over pairs of dots will be present and entanglement can still be manipulated due to spectral shifts $\sim g^2/\Delta$.

In order to verify the collapse of width of translational-symmetry-induced bands we numerically investigate the spectrum of a chain of four dots coupled via three cavities. We set $\omega_0/g = 10000$, $\Delta/g = 30$, and $\omega_e = 3\omega_t = 3 g$, as an example, which corresponds to a realistic excitonic frequencies and Zeeman splittings in self-assembled InAs/GaAs dots. We also truncate cavity modes to four states to perform exact diagonalization of the system. For these parameters $(g/\Delta)^4\sim 10^{-6}$. The energies of the first 2000 (out of 16384) states are shown in Fig.~\ref{fig:QDs}(d). In order to examine band splitting due to propagation of excitations we plot energy differences between the nearest energy states, i.e. $E_{n+1}-E_n$, in Fig.~\ref{fig:QDs}(e). Figure~\ref{fig:QDs}(f) shows the same energy differences as in Fig.~\ref{fig:QDs}(e), except we artificially restrict C-1 and C-3 modes to couple only to $\v{1}\lr\v{3}$ transitions and the C-2 mode to couple only to $\v{0}\lr\v{2}$ transitions in the adjacent dots. These constraints factorize the system into non-interacting segments, with one cavity mode per segment, and eliminate band splitting due to sequences of type (\ref{eq:QDs:split}). Comparison of the plots shows that the removed splittings (the upper highlighted area) are indeed in the range $\sim(g/\Delta)^4$. The bottom highlighted energy range falls below standard numerical diagonalization accuracy ($\sim 10^{-13}$ for matrices with $\O(1)$ entries). To confirm the $\sim(g/\Delta)^4$ splitting due sequence (\ref{eq:QDs:split}) further we can numerically identify eigenstates with the largest overlap with $\v{1010,a_1^\dag}$ and $\v{1010,a_3^\dag}$ states. The splitting between the corresponding energies is found to be $1.53648\times 10^{-5} g$ which is consistent with the above description. Further numerical confirmation require identification of states, and will be done for sub-systems of two and three dots below.

\subsubsection{Two-dots subsystem}
\label{sec:Systems:QDs:two}

In the intermediate resonance regime, each pair of dots develops specific symmetry breaking in relations (\ref{eq:QB:NI-Omega}). It has been found earlier\cite{solenov3} that in the system of two dots with only one $\v{0}\lr\v{2}$ transition used and for certain strength of interaction, Rabi frequencies for transitions that involve only one excitation, $\v{2i}\lr\v{0i}$ and $\v{i2}\lr\v{i0}$, are indistinguishable (local) for different $i=0,1$, i.e., $\Omega_{2i,0i}\to\Omega_{2,0;\text{at dot 1}}$ and $\Omega_{i2,i0}\to\Omega_{2,0;\text{at dot 2}}$. The two transitions  $\v{22}\lr\v{02}$ and $\v{22}\lr\v{20}$ are distinguishable from any of the $\v{2i}\lr\v{0i}$ and $\v{i2}\lr\v{i0}$ transitions respectively. Our system involves at least two more transitions per dot $\v{1}\lr\v{3}$ and $\v{1}\lr\v{2}$, which enable multiple other transitions involving two or multi-dot states. In order to obtain the symmetry relations between the corresponding Rabi frequencies we analyze the system numerically. 

\begin{figure}\begin{center}
\includegraphics[width=0.9\columnwidth]{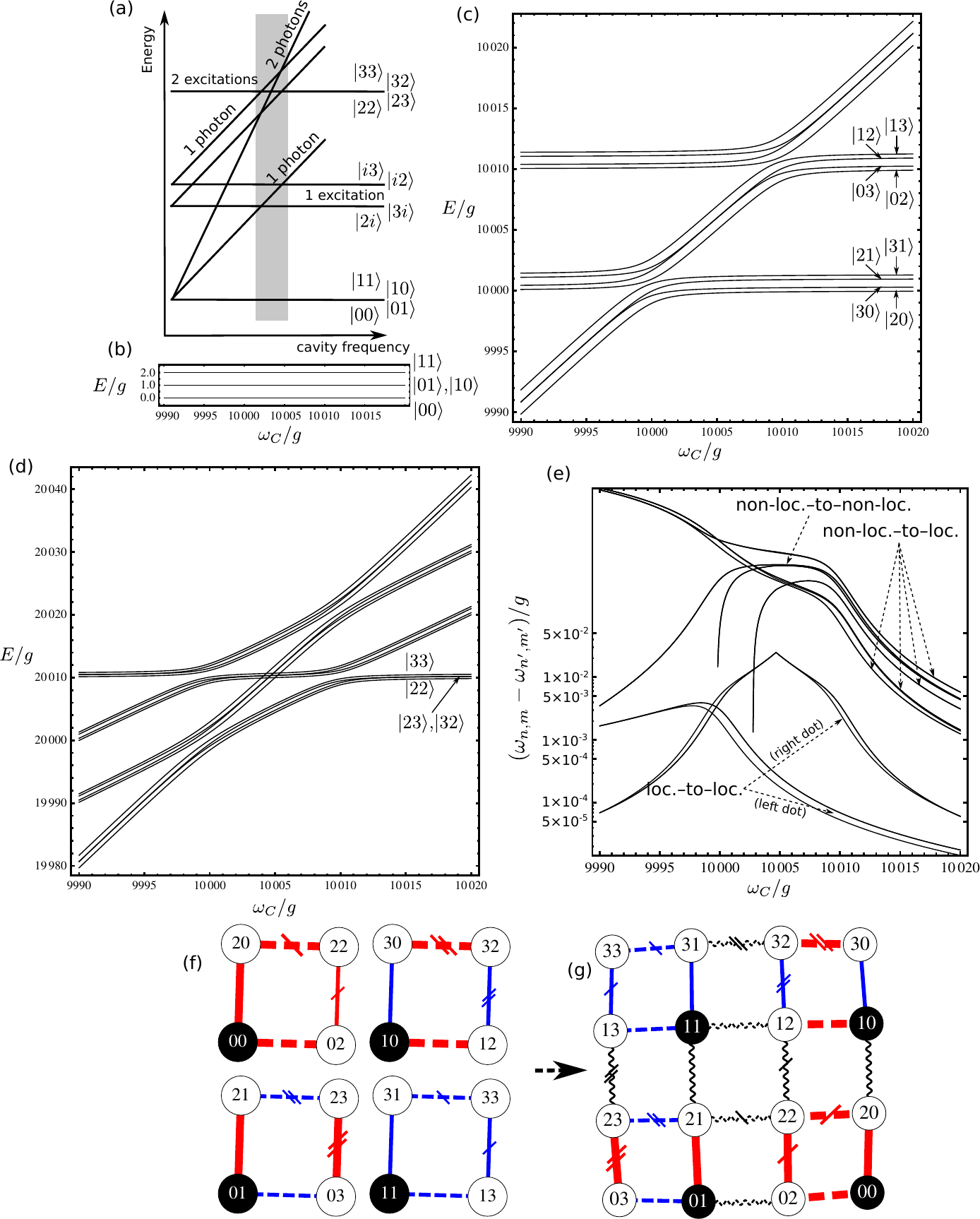}
\caption{\label{fig:QDs-2QB}
Energy spectrum of two quantum dots and the corresponding transition networks (graphs). 
(a) Total energy, schematically. Highlighted area shows location of anti-crossings of interest. (b-d) Numerically obtained parts of the spectrum for $\omega_0/g=10^4$, $\Delta/g=10$, $\omega_e=3\omega_t=g$. (e) Transition frequency differences that define reduction of symmetry (\ref{eq:QB:NI-Omega}). (f) A set of graphs outlining the symmetry of connections (Rabi frequencies), when only $\cal V$ transitions are used in pulse (\ref{eq:QW:V-gen}). (g) The symmetry of the network with both  $\cal V$ and $\cal H$ transitions (except for $\v{0}\lr\v{3}$) addressed by pulse (\ref{eq:QW:V-gen}). Non-local transitions are shown as single- or double-crossed lines. Lines of the same type mark transitions that are indistinguishable in the intermediate resonance regime. Other transitions are, in general, distinguishable. Transition $\v{0}\lr\v{3}$ adds ``periodic boundary conditions'' to graph (g) transforming it into a two-dimensional hyper-cycle\cite{Fedichkin1,Fedichkin2} graph (torus).
}\end{center}\end{figure}

We begin with the double-dot system, describing one segment of the register. In such segment, Fig.~\ref{fig:QDs}(b), quantum dots are coupled via a single cavity mode interacting with transitions $\v{0}\lr\v{2}$ and $\v{1}\lr\v{3}$ ($\cal V$ polarization only in this case). The schematic energy spectrum of the system as a function of the cavity mode frequency $\omega_C$ is shown in Fig.~\ref{fig:QDs-2QB}(a). The exact numerically obtain spectrum for $\Delta=10 g$, $\omega_0=10^4g$, $\omega_e=3\omega_t=g$ is shown in Fig.~\ref{fig:QDs-2QB}(b-d), where part (b) shows the qubit computational basis subspace energy range, part (c) shows states with one excitation and part (d) shows states with two excitations. The cavity frequency is varied in the range $\omega_0-\Delta \leq \omega_C \leq \omega_0+2\Delta$. In order to investigate interaction-induced symmetry reduction in the intermediate resonance regime, in Fig.~\ref{fig:QDs-2QB}(e) we plot numerically obtained transition frequency differences $\omega_{n,m}-\omega_{n',m'}$ as a function of $\omega_C$. We notice that all these differences fall into three categories: (i) local-to-local differences, (ii) non-local-to-local differences, and (iii) non-local-to-non-local differences. Group (i) has differences
\begin{eqnarray}\label{eq:QDs:group-i}
\omega_{20,00}-\omega_{21,01},\quad \omega_{30,00}-\omega_{31,01},\quad
\omega_{20,10}-\omega_{21,11},
\end{eqnarray}
and the other three with all dot indexes swapped. Note that the later three are larger because transitions are based on the right dot exciton, which is closer to the cavity spectrally for that cavity mode frequency range. Group (ii) has differences
\begin{eqnarray}\label{eq:QDs:group-ii}
\omega_{20,00}-\omega_{22,02},\quad \omega_{21,01}-\omega_{23,03},\quad \omega_{30,10}-\omega_{32,12},
\\\nonumber
\omega_{31,11}-\omega_{33,13},\quad \omega_{21,11}-\omega_{23,13},\quad \omega_{20,10}-\omega_{22,12}, 
\end{eqnarray}
and the other six with all dot indexes swapped. Group (iii) has differences 
\begin{eqnarray}\label{eq:QDs:group-iii}
\omega_{22,02}-\omega_{23,03},\quad \omega_{33,13}-\omega_{32,13},\quad
\omega_{22,12}-\omega_{23,13}
\end{eqnarray}
and the other three with all dot indexes swapped. In the (i) group all differences fall below $g\times 10^{-4}$ at $\omega_C \sim \omega_0+2\Delta$, i.e., at the right edge of the plotted frequency range. In groups (ii) and (iii) the values are at least two orders of magnitude larger, and the differences in group (iii) are of approximately the same magnitude as in group (ii). Therefore we can set the overall pulse profile $\Phi(t)$ to be sufficiently fast (broad band) to render transitions in group (i) indistinguishable, and yet sufficiently slow (narrow band) to distinguish transitions in groups (ii) and (iii), which defines the intermediate resonance regime. The symmetry of transitions is outlined in Figs.~\ref{fig:QDs-2QB}(f) and (g). In part (f) only $\v{0}\lr\v{2}$ and $\v{1}\lr\v{3}$-based transitions are shown and part (g) also has $\v{1}\lr\v{2}$-based transitions. Connecting lines of the same type mark indistinguishable transitions that can not be addressed independently by the corresponding resonant component of the multicolor pulse (\ref{eq:QW:V-gen}). The crossed lines denote different line types and correspond to transitions involving states with two excitations. Transitions of the same color become indistinguishable in the limit $g\to 0$, as required by relations (\ref{eq:QB:NI-Omega}). Finally, when transition $\v{0}\lr\v{3}$ is added, it appends ``periodic boundary conditions'' to graph (g) transforming it into a two-dimensional hyper-cycle\cite{Fedichkin1,Fedichkin2} graph (torus), i.e., nodes $\v{03}$ and $\v{33}$, $\v{01}$ and $\v{02}$, $\v{32}$ and $\v{33}$, etc., become connected.

\subsubsection{Three-dots subsystem}
\label{sec:Systems:QDs:three}

\begin{figure}\begin{center}
\includegraphics[width=0.99\columnwidth]{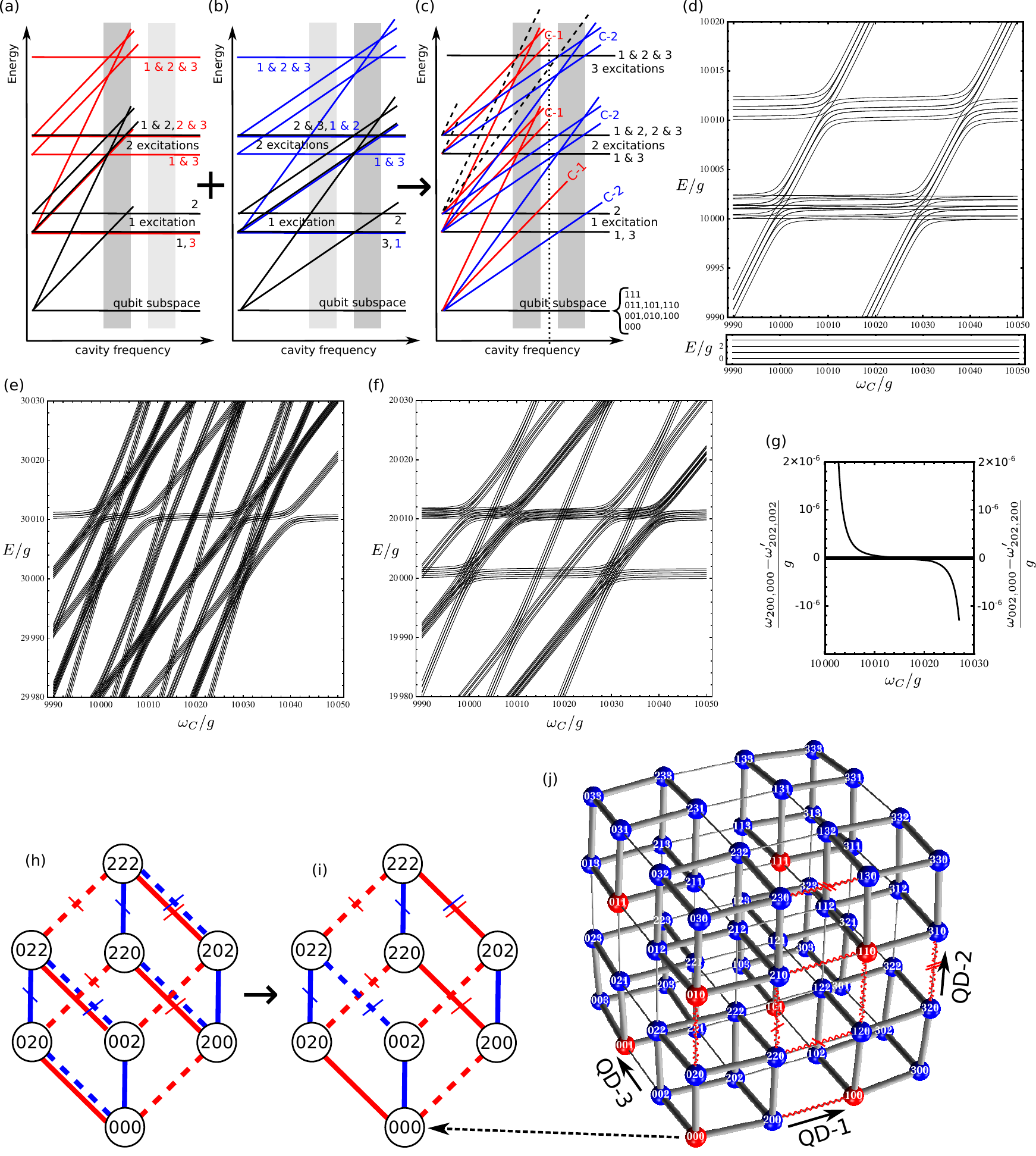}
\caption{\label{fig:QDs-3QB}
Energy spectrum of three quantum dots and the corresponding transition network. (a-c) Schematic representation of the energies as a function of the cavity C-1 frequency, provided the ratio of cavity mode frequencies remains constant.
(d-f) Numerically computed spectrum; parameters are the same as in Fig.~\ref{fig:QDs-2QB}. (g) Difference of two transition frequencies demonstrating vanishing effect of excitation propagation through the register. (h-j) Symmetry of transition network; lines of the same type correspond to indistinguishable transitions; see text for explanation. 
}\end{center}\end{figure}

In order to understand how symmetry (\ref{eq:QB:NI-Omega}) is broken in a larger segment of the linear chain register we must investigate a three-dot subsystem, as shown in Fig.~\ref{fig:QDs}(c). The spectrum of three quantum dots and two cavity modes is substantially more complex. Yet its schematic structure can be recovered through the following simple procedure outlined in 
Fig.~\ref{fig:QDs-3QB}(a-c). The double-dot spectrum of the left two dots interacting via cavity C-1 as a function of $\omega_\text{C-1}$ [black lines in Fig.~\ref{fig:QDs-3QB}(a)] is shifted up by the exciton energy in the third dot if the later is excited [red lines in Fig.~\ref{fig:QDs-3QB}(a)]. Similarly, the spectrum of the second and the third dots interacting via cavity mode C-2 [black lines in Fig.~\ref{fig:QDs-3QB}(b)] is shifted up if the first dot is excited [blue lines in Fig.~\ref{fig:QDs-3QB}(b)]; still as a function of $\omega_\text{C-1}$ but with $\omega_\text{C-1}/\omega_\text{C-2} = {\rm const}$. In both cases a series of anti-crossings develop where bands intersect. The superposition of part (a) and part (b) gives the schematic structure of the spectrum of the three-dot system shown in Fig.~\ref{fig:QDs-3QB}(c). Note that not all intersections lead to anti-crossings. Many states are orthogonal and, hence, can not couple. Note also that states with higher photon count (some of which are shown by dashed lines) do not interfere appreciatively with the shown states in (and in between) the shaded regions. For example, two two-photon lines originating from the one-excitation line of the second dot (shown as dashed) can anti-cross with the one-photon lines in the three-excitation region of the spectrum. This process, however, involves transferring excitations between dots 1 and 3, which is a $\sim (g/\Delta)^4$ process as discussed above, and the resulting splitting can be neglected. We obtain the exact spectrum numerically [see Fig.~\ref{fig:QDs-3QB}(d-f)] for the same parameters as used in the two-dot case above (shown in Fig.~\ref{fig:QDs-2QB}). The ratio between cavity frequencies was set to
\begin{eqnarray}\label{eq:QDs:C-1-2}
\frac{\omega_\text{C-1}}{\omega_\text{C-2}} = 
\frac{\omega_0 + 2\Delta}{\omega_0 - \Delta}
\end{eqnarray}
such that at the middle point, marked by the vertical dashed line in Fig.~\ref{fig:QDs-3QB}(c), one cavity is above the top single-excitation band by $\Delta$ and the other is below the bottom single-excitation lines by $\Delta$, as suggested earlier. The two- and three-excitation parts of the spectrum in Fig.~\ref{fig:QDs-3QB}(e) and (f) have lower-excitation parts with additional photons superimposed on them, making them hard to read. This does not change the simple anti-crossing structure schematically shown in Fig.~\ref{fig:QDs-3QB}(c) because these overlapped bands do not interact in the cavity frequency region of interest. As before, the resonators were modeled using four states. To further illustrate that similar transitions that belong to different segments of the register remain unaffected by each other, we plot energy difference $\omega_{200,000} - \omega_{202,002}$ in Fig.~\ref{fig:QDs-3QB}(g). It remains at, or below, $g\times 10^{-6}$ level for the cavity mode frequencies of interest, which indicates that transition $\v{0}\lr\v{2}$ in the first dot is unaffected by the similar transition in the third dot in the intermediate resonance regime. The symmetry of transitions is outlined in Fig.~\ref{fig:QDs-3QB}(h-j). Figures~\ref{fig:QDs-3QB}(h-i) illustrate how the symmetry is obtained for each subgraph using the example of a subgraph based on state $\v{000}$. Figure~\ref{fig:QDs-3QB}(j) shows the entire network of transitions. Specifically, in Fig.~\ref{fig:QDs-3QB}(h) red lines correspond to breaking of symmetry (\ref{eq:QB:NI-Omega}) due to QD-1 and QD-2 double-dot system and blue lines correspond to QD-2 and QD-3 double-dot system, with the same graphic notation as in Fig.~\ref{fig:QDs-2QB}(f). For example, the frequency of transition $\v{202}\lr\v{222}$ is non-negligibly shifted by both the first (red) and the second (blue) double-dot systems, making it distinguishable from both $\v{022}\lr\v{002}$ and $\v{220}\lr\v{200}$ as shown in Fig.~\ref{fig:QDs-3QB}(i). On the other hand, transitions remain indistinguishable in each pair: \{$\v{000}\lr\v{200}$ and $\v{002}\lr\v{202}$\}, \{$\v{000}\lr\v{002}$ and $\v{200}\lr\v{202}$\},
\{$\v{020}\lr\v{220}$ and $\v{022}\lr\v{202}$\}, \{$\v{020}\lr\v{022}$ and $\v{220}\lr\v{222}$\}. The frequency difference corresponding to the first pair is shown in Fig.~\ref{fig:QDs-3QB}(g). This is consistent with suppression of excitonic propagation by one segment along the chain as discussed above. It can also be numerically verified that transitions involving the middle dot $\v{i0j}\lr\v{i2j}$ with different combinations of $i,j$ are distinguishable (have substantially larger frequency differences). Note that accidental degeneracies in the transition network can still render some transition indistinguishable at some specific values of the system parameters, including $\omega_\text{C-$n$}$ and $\Delta$. The front face of the cube in Fig.~\ref{fig:QDs-3QB}(j) is the cross-section representing a two-dot subsystem shown in Fig.~\ref{fig:QDs-2QB}(g) when the third qubit is in state $0$. The wavy and broken lines show symmetry of $\v{1}\lr\v{2}$ transitions in this cross-section. It is not shown on other parts of the cubic lattice to avoid clutter. If $\v{0}\lr\v{3}$ based transitions are also accounted for, graph (j) closes into a three-dimensional hyper-cycle\cite{Fedichkin1,Fedichkin1} graph, i.e., into a 3D crystal lattice with periodic boundary conditions and the primitive cell defined by graph (j).

Each additional qubit will increase the dimension of the transition network grid by one. The $N$-dot chain, therefore, creates a base-4 $N$-dimensional hypercube graph (or hyper-cycle graph if all four transitions per dot are accounted for) with structured network of local and non-local transitions. The symmetry of transitions in such network can be derived following the same procedure as outlined in Fig.~\ref{fig:QDs-3QB}(h-g), keeping in mind that transitions in quantum dots separated by more then one dot do not affect each other in the intermediate resonance regime. Lower-dimensional cross-sections can be considered to construct entangling or non-entangling gates involving the desired number of qubits. The procedure of constructing quantum gates and examples involving some of these lower-dimensional cube graphs are discussed in the next sections. The cavity-based connections discussed above and shown in Fig.~\ref{fig:QDs} are not the only possible scalable arrangement. It is also possible, e.g., to couple cavity modes with one parity (of the index) to $\cal H$ transitions and cavity modes with the other parity to $\cal V$ transitions in a similar chain. This will also remove spectral crowding in the intermediate resonance regime, but it will create a different network of transitions. The network of this type is discussed in Subsection \ref{sec:Systems:Transmons}, where it is the most natural option.

\subsection{Defects in diamond}
\label{sec:Systems:Diamond}

\begin{figure}\begin{center}
\includegraphics[width=0.85\columnwidth]{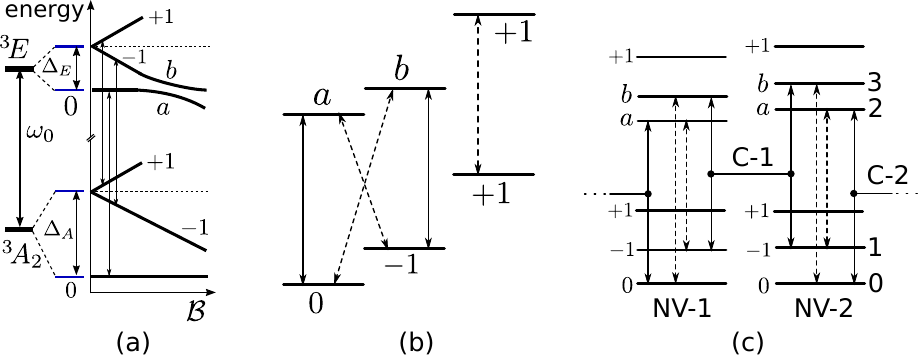}
\caption{\label{fig:NVs}
Level diagram for nitrogen vacancy (NV) centers in diamond.
(a) Relevant energy levels of a NV-center\cite{solenov2,solenov3,Acosta,Doherty} as a function of magnetic field $\cal B$. (b) Energy levels in the magnetic field mixing states in the upper triplet. Qubits are encoded by states $\v{0}$ and $\v{-1}$ in each NV center. The energy level diagram is similar to that of a quantum dot shown in Fig.~\ref{fig:QDs}(a). (c) A single element of a chain of NV centers (quantum register) connected via different cavity modes similarly to quantum dot system shown in Fig.~\ref{fig:QDs}(b).
}\end{center}\end{figure}

Defects in diamond have six optically addressable states shown schematically in Fig.~\ref{fig:NVs}(a). In each triplet, the dublet is split off from the spin-0 state due to crystal strain around the defect.\cite{solenov2,solenov3,Acosta,Doherty} Each dublet has two spin states and can be split with the magnetic field.\cite{Togan,Yale} Optical transitions conserve spin in this system. However, at sufficiently strong magnetic fields, the lowest two states of the higher energy triplet mix, allowing for the ``cross'' transitions. In this case the spectrum becomes similar to that of a self-assembled quantum dot, c.f., Fig.~\ref{fig:QDs}(a) and Fig.~\ref{fig:NVs}(b), except for the polarization dependence. As the result, optical control in the defect centers can be performed in the same fashion.\cite{solenov3}

As in the case of quantum dots, the propagation of excitations by one segment [see Fig.~\ref{fig:NVs}(c)] along the chain involves four off-resonance absorptions or emissions of cavity mode photons, each contributing a factor of $\sim g/\Delta$ if frequencies of transitions and cavity modes are arranged the same way as in the previous subsection. The translation-induced energy bands will have widths of $\sim g(g/\Delta)^4$, which are spectrally indistinguishable in the intermediate resonance regime. The corresponding states will, therefore, remain effectively local, and the system will split into pairs of defects that can be locally addressed by the multicolor control pulses. The pulse will temporarily create graphs of types shown in Figs.~\ref{fig:QDs-2QB} and \ref{fig:QDs-3QB}, performing continuous time quantum walks in effective time $\tau$ with the desired outcome as discussed in the next section.

\subsection{Superconducting transmon qubits}
\label{sec:Systems:Transmons}

\begin{figure}\begin{center}
\includegraphics[width=0.85\columnwidth]{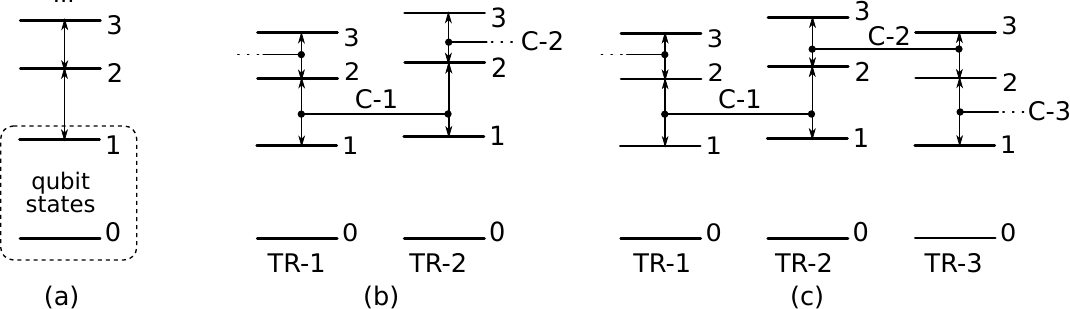}
\caption{\label{fig:TRs}
Level diagrams for superconducting transmon architecture. (a) The first four energy levels of a single transmon system. The qubit is encoded by states $\v{0}$ and $\v{1}$. (b) Subsection of a multiqubit transmon register consisting of two spectrally distinct transmons connected via a cavity mode. (c) Connection to the next transmon along the chain. All cavity modes are orthogonal to each other.
}\end{center}\end{figure}

Superconducting transmon qubit systems are substantially different from the systems described in the two previous examples. A transmon is a variation of a cooper-pair box qubit in which Josephson energy, $E_J$, dominates over the charging energy.\cite{Koch,Devoret,Paik} In this limit the system resembles a heavy quantum particle in a periodic $-E_J\cos\phi$ potential subject to periodic boundary condition on phase $\phi$ of the superconducting order parameter. The low-energy spectrum is approximately harmonic\cite{Koch}
\begin{eqnarray}\label{eq:TRs:Es}
E_n = \left( \omega_{01} - \frac{\alpha}{2} \right) n
+ \frac{\alpha}{2} n^2,
\end{eqnarray}
with small negative anharmonicity $\alpha$ defined as 
\begin{eqnarray}\label{eq:TRs:alpha}
\alpha = \omega_{01} - \omega_{12}
\end{eqnarray}
where $\omega_{ij} = E_j-E_i$.
The first four energy levels are shown in Fig.~\ref{fig:TRs}(a) schematically. In order to correctly represent the spectrum at higher energies or at large anharmonicities, Eq.~(\ref{eq:TRs:Es}) must be adjusted\cite{Koch} to include tunneling due to periodic boundary conditions on $\phi$ and the correct shape of the Josephson potential energy as a function of $\phi$. Transmons are designed\cite{Koch} to have $|\alpha/\omega_{01}|$ below $0.1$ with $\alpha/\omega_{01}\sim -0.01$ for low noise transmons.\cite{Paik} In these systems microwave field can strongly couple to consecutive transitions, i.e., $\v{0}\lr\v{1}$, $\v{1}\lr\v{2}$, $\v{2}\lr\v{3}$ etc, and nearly harmonic approximation (\ref{eq:TRs:Es}) is sufficient.

A chain of interacting transmons can be organized by coupling adjacent transmons via microwave cavity modes. In order to attenuate the propagation of excitations through the chain to $O([g/\Delta]^4)$ as before, we must design cavity modes such that the corresponding frequencies are detuned by $\Delta$ to the red and to the blue from $\v{2}\lr\v{3}$ and $\v{1}\lr\v{2}$ transition frequencies respectively, alternating through the chain. The chain can be approximately or exactly base-two translationally symmetric. A single element of the chain is shown in Fig.~\ref{fig:TRs}(b), and connection to the next segment is shown in Fig.~\ref{fig:TRs}(c). 

When transition frequencies $\omega_{12}$ and $\omega_{23}$ are detuned from the same respective transitions in the adjacent transmon by $\sim\Delta$ with $g/\Delta \ll 1$ such that energy gaps $\sim g^2/\Delta$ are resolvable by microwave pulses and gaps $\sim g^4/\Delta^3$ are not resolvable, each pair of transmons is in the intermediate resonance regime described in Ref.~\onlinecite{solenov3}. The energy cost for excitation to propagate from one pair to the next symmetrically equivalent pair is $\sim g(g/\Delta)^4$, e.g.,
\begin{eqnarray}\label{eq:TRs:split}
\v{i21j..,a_1^\dag}
\xrightarrow[]{\sim g/\Delta}
\v{i31j..}
\xrightarrow[]{\sim g/\Delta}
\v{i21j..,a_2^\dag}
\xrightarrow[]{\sim g/\Delta}
\v{i22j..}
\xrightarrow[]{\sim g/\Delta}
\v{i21j..,a_3^\dag}
\end{eqnarray}
Therefore, base-2 translation-induced energy shifts will be indistinguishable to the control pulse, which eliminates spectral crowding, as discussed in the previous subsections.

\begin{figure}\begin{center}
\includegraphics[width=0.99\textwidth]{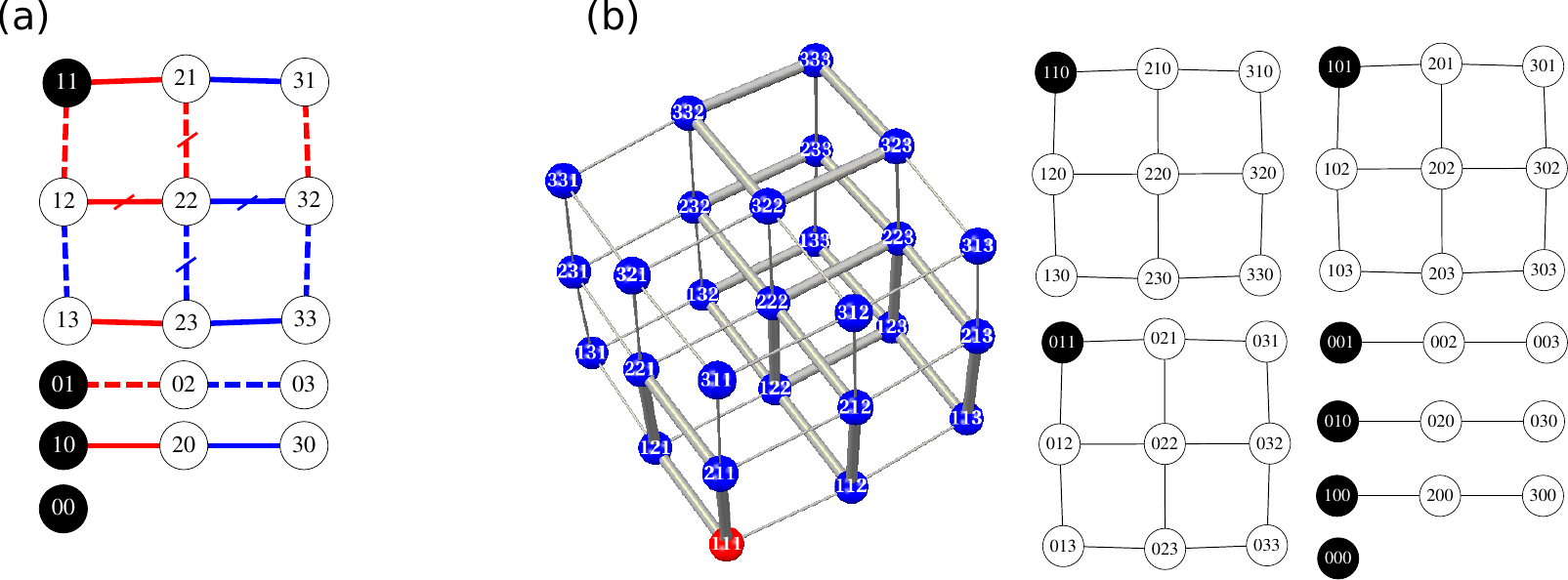}
\caption{\label{fig:TRs-3QB}
A set of graphs representing (a) two- and (b) three-transmon subsection of the chain register. The graphs are disjointed in both cases because transitions skipping one energy level, e.g., $\v{1}\lr\v{3}$, are not practically accessible. The symmetry of transition network is explicitly shown for two-transmon segment in part (a). The $\v{1}\lr\v{2}$ and $\v{2}\lr\v{3}$ subsets involving transmons 1 and 2, and transmons 2 and 3 respectively are highlighted in the $\v{111}$-based graph in part (b) to show structure.
}\end{center}\end{figure}

The network of transitions available to pulse-induced quantum walks differ from the one shown in Figs.~\ref{fig:QDs-2QB} and \ref{fig:QDs-3QB} because only three states are involved. Accessible graphs describing a sub-system of two transmons are shown in Fig.~\ref{fig:TRs-3QB}(a). The graphs are disjointed because transitions that skip one state are not available. The symmetry of transitions is shown on the same plot. It is deduced by observing the that levels participating in transitions such as $\v{31}\lr\v{32}$ involve the same arrangement of anti-crossings as $\v{11}\lr\v{12}$ (without two-photon line). Levels participating in transitions of type $\v{33}\lr\v{23}$ have single-photon line at the bottom (near $\v{23}$), which make them similar to effectively local $\v{11}\lr\v{21}$ transitions, except for lower transition frequency, $\omega_{23}$. Transitions such as $\v{32}\lr\v{22}$ should be distinct from $\v{33}\lr\v{23}$ because participating energy levels involve two-photon line near $\v{22}$. The set of graphs accessible for a three qubit subsystem are shown in Fig.~\ref{fig:TRs-3QB}(b). The transition network symmetries can be obtained by procedure similar to the one outlined in Fig.~\ref{fig:QDs-3QB}(h-i). Note that, due to its nearly harmonic spectru,m transmon systems can be affected by accidental degeneracies (and anti-crossings) more substantially than systems of quantum dots. This, however, does not invalidate the intermediate resonance regime approach because cavity-transmon coupling strength $g$ (and bandwidths of the pulses) is typically much smaller then anharmonicity $\alpha$, even though the latter is much smaller than $\omega_{01}$ in each transmon. In general, the largest graph is based on state $\v{1..1}$ and resembles a hyper-cube lattice of three nodes in each dimension. All other graphs are cross-sections of that graph with one or several $\v{0}$ states in place of $\v{1}$. Note that, as described earlier, the ``non-interacting'' state labels refer to states that can be non-local, but are connected to those non-interacting states adiabatically when $g\to 0$.

Finally we note that base-three hypercube networks of type shown in Fig.~\ref{fig:TRs-3QB} can also appear in system of quantum dots when odd and even-parity cavity modes are coupled to transitions with different polarizations. In that case states $\v{1}$, $\v{2}$, and $\v{3}$, can be mapped, e.g., onto states $\v{\uparrow}$, $\v{\Uparrow}$, and $\v{\downarrow}$ respectively. At the same time, in this case cube graphs will involve more than one computational basis state. Therefore quantum walks designed to perform certain gates based on graphs representing transmon architecture will not be necessarily portable to quantum dots architecture with orthogonally polarized cavity modes.

\section{Quantum gates via quantum walks}
\label{sec:QB}

In this section we discuss structure of $\Lambda$ necessary to implement entangling and local (single-qubit) quantum gates and give several examples of such implementations.
In what follows we will focus primarily on the  reduction of symmetry (\ref{eq:QB:NI-Omega}) based on the intermediate resonance regime and the cavity-mediated interactions discussed in the previous section. We will demonstrate how one-, two-, and three-dimensional cross sections (sub-graphs) of the multidimensional graphs corresponding to the scalable qubit register can be used to perform local and entangling operations. When degeneracy (\ref{eq:QB:NI-Omega}) is lifted differently, the gates can be constructed in a similar fashion, but different graphs, and, hence, pulse spectra, might be necessary in each case. Furthermore, because violation of symmetry (\ref{eq:QB:NI-Omega}) is a manifestation of physical interactions between qubits, some entangling gates might not be accessible in certain cases. This is not surprising because necessary physical interactions might simply be absent.

\subsection{Single-qubit quantum gates}\label{sec:1QB}

Single-qubit quantum gates in systems with actively used auxiliary states are the simplest examples of $P\Lambda P\neq\Lambda$ gates implemented via quantum walks. Here we give few examples of gates, some of which are performed routinely in different quantum computing systems,\cite{Wrachtrup,Awschalom,Kennedy,Hanson} to demonstrate their connection with a (more general) quantum-walks-based approach investigated in this paper. 

The first example is Z gate.\cite{nielsenchuang} This gate flips the sign of the amplitude for one of the qubit's state, i.e.,
\begin{eqnarray}\label{eq:1QB:UZ}
U_g({\rm Z}) = \sigma_z \equiv \left(
\begin{array}{cc}
1 & 0 \\
0 & -1
\end{array}
\right)
\end{eqnarray}
In the simplest case, a single auxiliary state is sufficient and we can choose the graph with the following adjacency matrix
\begin{eqnarray}\label{eq:1QB:Lambda-Z}
\Lambda = \left(
\begin{array}{ccc}
0 & 0 & \Omega \\
0 & 0 & 0 \\
\Omega^* & 0 & 0
\end{array}
\right)
\end{eqnarray}
in the basis $\{\v{2},\v{1},\v{0}\}$, i.e. transition between states $\v{0}$ and $\v{2}$ is addressed (activated) by external pulse with Rabi frequency $\Omega$. Upon examination of the solution of this effectively two-state problem (see Sec~\ref{sec:LG:CH2}) it is evident that Eq.~(\ref{eq:1QB:UZ}) is obtained from Eqs. (\ref{eq:QW:Ug}), (\ref{eq:QW:evolve}), and (\ref{eq:QW:walk}) when the walk is terminated at $\tau = (2n+1)\pi/|\Omega|$, where $n$ is any (non-negative) integer.

Another example is a (single-qubit) swap gate with arbitrary phase change, i.e.,
\begin{eqnarray}\label{eq:1QB:U-swap}
U_g(swap,\phi) = \left(
\begin{array}{cc}
0 & e^{i\phi} \\
e^{-i\phi} & 0
\end{array}
\right)
=\sigma_x\cos\phi - \sigma_y\sin\phi
\end{eqnarray}
Using the same three states as before, one of which is an auxiliary state, we can set the graph to have adjacency matrix
\begin{eqnarray}\label{eq:1QB:Lambda-swap}
\Lambda = \left(
\begin{array}{ccc}
0 & |\Omega_1|e^{i\varphi_1} & 0 \\
|\Omega_1|e^{-i\varphi_1} & 0 & |\Omega_2|e^{-i\varphi_2} \\
0 & |\Omega_2|e^{i\varphi_2} & 0
\end{array}
\right)
\end{eqnarray}
in the basis $\{\v{1},\v{2},\v{0}\}$. Examination of quantum walks on such graph (chain of three states, see Sec.~\ref{sec:LG:CH3}) shows that if we set $|\Omega_1| = |\Omega_2|$ and $\varphi_1 - \varphi_2 = \phi$, gate (\ref{eq:1QB:U-swap}) is obtained provided the walk is terminated at time $\tau = (2n+1)\pi /\sqrt{2|\Omega_1|^2}$, where $n$ is any (non-negative) integer.

Finally, we consider an example of implementing the Hadamard gate
\begin{eqnarray}\label{eq:2QB:CZ}
{\rm H} =
\frac{1}{\sqrt{2}}\left(
\begin{array}{cc}
1 & 1 \\
1 & -1
\end{array}\right)
\end{eqnarray}
which is widely used in algorithms and error correction codes.\cite{nielsenchuang} Similarly to the previous example, it can be performed via a quantum walk on the graph with adjacency matrix (\ref{eq:1QB:Lambda-swap}). In this case (see Sec.~\ref{sec:LG:CH3}) we must set $\phi_1-\phi_2=\pi$ and $|\Omega_2|=|\Omega_1|/(\sqrt{2}-1)$. Hadamard gate evolution operator (\ref{eq:1QB:U-swap}) is obtained when the walk is terminated at time $\tau = (2n+1)\pi /\sqrt{|\Omega_1|^2+|\Omega_2|^2}$, where $n$ is any (non-negative) integer.

Note that in all three cases, Eq.~(\ref{eq:QW:QLP}) is satisfied and the probability is completely returned back to the qubit nodes ($\v{0}$ and $\v{1}$) at time $\tau$. While such abrupt termination of the walk may seem unnatural, we should note that $\tau$ is not the physical time in the system. It is the overall integral magnitude of the external control field [see Eq.~(\ref{eq:QW:tau})], which can be controlled with high accuracy in experiment. The change of the control field with real physical time is typically a smooth function with maximum at (physical) time $t=(t_1+t_2)/2$ and with sufficiently small values at $t_1$ and $t_2$, e.g., $\Phi(t)\sim\exp\{-\sigma^2[t-(t_1+t_2)/2]^2\}$.

\subsection{Two-qubit quantum gates}\label{sec:2QB}

One of the most important two-qubit entangling gates is the CNOT (Control-NOT) gate.\cite{nielsenchuang} It is defined as
\begin{eqnarray}\label{eq:2QB:CNOT}
{\rm CNOT} =
\left(
\begin{array}{cccc}
1 & 0 & 0 & 0 \\
0 & 1 & 0 & 0 \\
0 & 0 & 0 & 1 \\
0 & 0 & 1 & 0
\end{array}
\right) = (I\otimes {\rm H}) {\rm CZ} (I\otimes {\rm H})
\end{eqnarray}
in the basis of, e.g., $\{\v{00},\v{01},\v{10},\v{11}\}$. It can be represented via two local Hadamard gates acting on one of the qubits and CZ (Control-Z) gate
\begin{eqnarray}\label{eq:2QB:CZ}
{\rm CZ} =
\left(
\begin{array}{cccc}
1 & 0 & 0 & 0 \\
0 & 1 & 0 & 0 \\
0 & 0 & 1 & 0 \\
0 & 0 & 0 & -1
\end{array}\right)
\end{eqnarray}
The CZ gate has a simple structure: it requires a set of {\it return} walks with the adjacency matrix restricted to
\begin{eqnarray}\label{eq:2QB:walk}
\v{\bf i'}\iv{\bf i'}e^{-i\tau\Lambda}\v{\bf i}\iv{\bf i} = 0,
\quad\quad
{\bf i}\neq{\bf i'}
\end{eqnarray}
Moreover, for the version of CZ given in Eq.~(\ref{eq:2QB:CZ}), the quantum walks, terminated at time $\tau$, must yield
\begin{eqnarray}\label{eq:2QB:walk-00}
e^{-i\tau\Lambda}\v{00} &=& \v{00},\\
\label{eq:2QB:walk-01}
e^{-i\tau\Lambda}\v{01} &=& \v{01},\\
\label{eq:2QB:walk-10}
e^{-i\tau\Lambda}\v{10} &=& \v{10},\\
\label{eq:2QB:walk-11}
e^{-i\tau\Lambda}\v{11} &=& -\v{11}
\end{eqnarray}
This is most easily achieved if the graph, corresponding to $\Lambda$, is separable into four disconnected subgraphs, each containing one of the two-qubit basis states, and each performing a {\it return} quantum walk when terminated at exactly the same time $\tau$. Only one subgraph must implement a non-trivial return walk. Other subgraphs are only required to produce a trivial return walk (effectively no evolution). 

\subsubsection{Adjacent qubits}\label{sec:2QB-near}

\begin{figure}\begin{center}
\includegraphics[width=0.5\columnwidth]{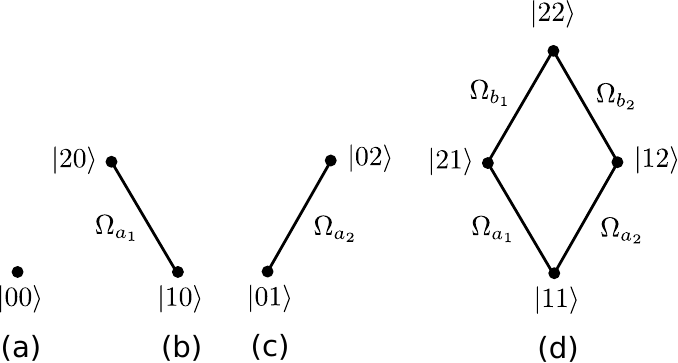}
\caption{\label{fig:2QB}
A set of graphs representing a two-qubit system with one active auxiliary state in each qubit and only one allowed transition, $\v{1}\leftrightarrow\v{2}$, in each qubit system. 
}\end{center}\end{figure}

As an example, consider graph in Fig.~\ref{fig:TRs-3QB}(a) with only $\v{1}\lr\v{2}$ transitions addressed by the pulse. Two-qubit CZ gates in adjacent qubits with graphs of type shown in Fig.~\ref{fig:QDs-2QB}(h) are similar up to renaming of vertices.
In this example only a single auxiliary state $\v{2}$ in each qubit system is set to interact with analogous state in the other qubit system. In the intermediate resonance regime described in the previous section, symmetry (\ref{eq:QB:NI-Omega}) is reduced such that we can set $\Omega_{11,21}\neq\Omega_{12,22}$ and $\Omega_{11,12}\neq\Omega_{21,22}$. We obtain a disconnected set of graphs shown in Fig.~\ref{fig:2QB}.
In this case Eq.~(\ref{eq:2QB:walk-00}) describes a walk on the trivial single-node graph (no evolution); Eqs.~(\ref{eq:2QB:walk-01}) and (\ref{eq:2QB:walk-10}) describe walks on two-state graphs (see Sec.~\ref{sec:LG:CH2}); and Eq.~(\ref{eq:2QB:walk-11}) involves a walk on a four-state square graph (see Sec.~\ref{sec:SQ}). 

A set of complex hopping amplitudes (edges) that satisfy Eqs.~(\ref{eq:2QB:walk-00}-\ref{eq:2QB:walk-11}) is not unique: an infinite number of solutions is possible. To demonstrate this we, first, define a dimensionless Rabi frequency $\xi$ as
\begin{eqnarray}\label{eq:2QB:Omega_to_a}
\xi = \Omega_\xi \tau / \pi
\end{eqnarray}
for every edge, where $\xi$ is $a_1$, $a_2$, $b_1$, or $b_2$ in this case. This makes all walks propagate over the same time interval $\tau$. We, then, set
\begin{eqnarray}\label{eq:2QB:n1n2}
|a_1| = n_1,
\quad\quad
|a_2| = n_2
\end{eqnarray}
to be positive even integers. This results in trivial return [see Eq.~(\ref{eq:LG:CH2:R_0}) in Sec.~\ref{sec:LG:CH2}] for all walks that start from states $\v{01}$ and $\v{10}$, thus satisfying Eqs.~(\ref{eq:2QB:walk-01}) and (\ref{eq:2QB:walk-10}).
Continuous time return walk through the square graph that contains state $\v{11}$ is investigated in Sec.~\ref{sec:SQ}. The absolute values of hopping amplitudes for the two bottom edges of this graph are already defined above. We have freedom to adjust the remaining two complex amplitudes, $b_1$ and $b_2$, and two phases, $\arg a_1$ and $\arg a_2$. As demonstrated in Sec.~\ref{sec:SQ}, a return walk on a square graph can be mapped onto a walk on a linear chain graph of four states (Sec.~\ref{sec:LG:CH4}). The latter allows for both trivial and non-trivial return walks [see Eqs.~(\ref{eq:LG:CH4:R_0}) and (\ref{eq:LG:CH4:R_pi}) in Sec.~\ref{sec:SQ}]. A non-trivial solution that satisfy Eq.~(\ref{eq:2QB:walk-11}) is parameterized by two odd integers $m$ and $n$. Without loss of generality we can set $0<m<n$. In this case the hopping amplitudes in graph~\ref{fig:2QB}(d) are bounded by condition
\begin{eqnarray}\label{eq:2QB:nm-limit}
m \le &\sqrt{n_1^2 + n_2^2}& \le n
\end{eqnarray}
and the solution is found from
\begin{eqnarray}\label{eq:2QB:sol}
\left\{
\begin{array}{l}
|a| \equiv \frac{|a_2b^*_1 - a_1b_2^*|}{
\sqrt{n_1^2 + n_2^2}}
= 
\frac{nm}{\sqrt{n_1^2 + n_2^2}}
\\
\frac{|a_1b_1 + a_2b_2|}{
\sqrt{n_1^2 + n_2^2}
} = 
\sqrt{
(n+m)^2 - (\frac{nm}{|a|}+|a|)^2
}
\end{array}
\right.
\end{eqnarray}

One specific example can be derived if we set $m=1$, $n=3$, $n_1=n_2=2$, and assume no complex phases for $a_1$ and $a_2$. In this case 
\begin{eqnarray}\label{eq:2QB:sol-123}
\left\{
\begin{array}{l}
|b_1-b_2| = 3/2
\\
|b_1+b_2| = \sqrt{7}/2
\end{array}
\right.
\to
\left\{
\begin{array}{l}
b_1 = \frac{\sqrt{7}e^{i\phi_i}+3e^{i\phi_{ii}}}{4}
\\
b_2 = \frac{\sqrt{7}e^{i\phi_i}-3e^{i\phi_{ii}}}{4}
\end{array}
\right.
\end{eqnarray}
where $\phi_i$ and $\phi_{ii}$ are two arbitrary real numbers. In this example all available transitions are activated to produce hopping amplitudes $a_1$, $a_2$, $b_1$, and $b_2$ given by Eqs.~(\ref{eq:2QB:n1n2}) and (\ref{eq:2QB:sol-123}). This, however, is not a necessary condition.

As another example, we can set $n_2=0$ (do not activate $a_2$ transition, see Fig.~\ref{fig:2QB}) and set $n_1$ to be an {\it odd} positive integer. This defines a non-trivial return walk for state $\v{10}$ instead of $\v{11}$ [the standard CZ gate (\ref{eq:2QB:CZ}) is recovered if we apply single-qubit Z gate to the first qubit]. The graphs with $\v{00}$ and $\v{01}$ states are now trivial one-node graphs. The graph that has $\v{11}$ node is now a linear chain graph with four nodes (see Sec.~\ref{sec:LG:CH4}), which is a subgraph of the square graph discussed above. The walk starting at $\v{11}$ must be a trivial return walk---parameters $m$ and $n$ must be non-equal {\it even} integers. Because $n_1$ is odd, it can always be chosen between $m$ and $n$ to satisfy Eq.~(\ref{eq:2QB:nm-limit}). As an illustration, we chose $m=2$, $n_1=3$, and $n=4$. From Eq.~(\ref{eq:2QB:sol}) we obtain $|b_1| = \sqrt{35}/3 \approx 1.972$ and $|b_2| = 8/3 \approx 2.667$. In this case the phases of all hopping amplitudes can be arbitrary.

\subsubsection{Next nearest neighbor qubits}\label{sec:2QB-far}

\begin{figure}\begin{center}
\includegraphics[width=0.99\columnwidth]{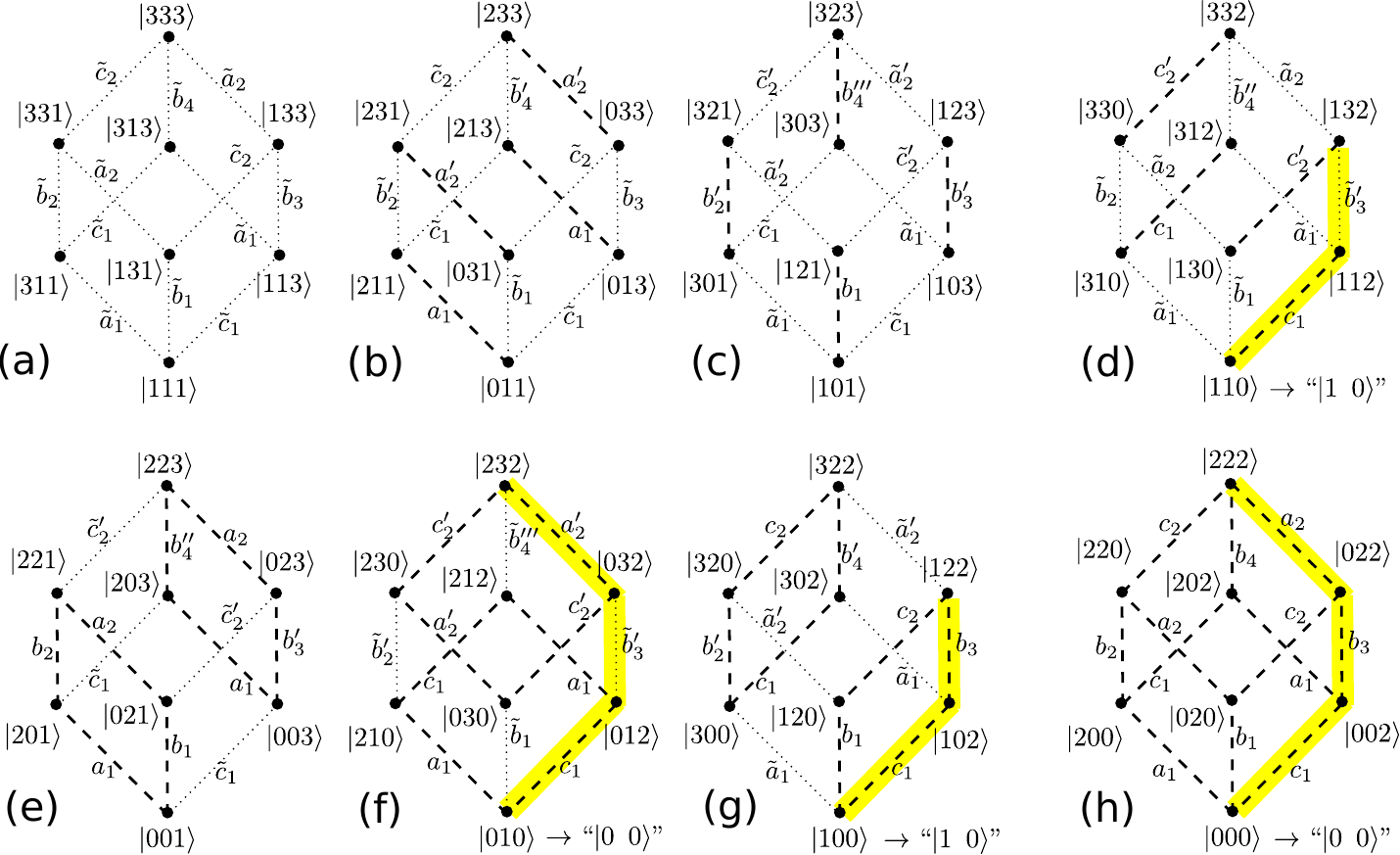}
\caption{\label{fig:2QB-far}
A set of graphs representing a sub-system of three quantum dots with  transitions $\v{0}\lr\v{2}$ (dashed lines) and $\v{1}\lr\v{3}$ (doted lines) allowed in each dot. Symmetry of transitions is shown via dimensionless Rabi frequencies. Subgraphs activated by the single multi-color pulse performing CZ gate on the first and the last qubits are highlighted (in yellow).
}\end{center}\end{figure}

Here we give another example of a CZ gate for qubits that are one qubit away from each other in the chain register discussed in the previous section. We will focus on graph
Fig.~\ref{fig:QDs-3QB}(j) that appear in, e.g., chains of quantum dots, and will avoid transitions based on $\v{1}\lr\v{2}$ and $\v{0}\lr\v{3}$ transitions in each dot. In this case the middle dot (QD-2) becomes part of the medium to carry interaction between the left and right dots. All available graphs are shown in Fig.~\ref{fig:2QB-far}.

We chose a single multi-color pulse approach as before, and activate five distinct transitions $a_2$, $b_3$, $c_1$, $a'_2 = a_2$, and $\tilde b'_3 = b_3$, where the dimensionless Rabi frequencies are defined by Eq.~(\ref{eq:2QB:Omega_to_a}) as before. The last two dimensionless frequencies are set equal to the first two to make quantum walks starting from nodes $\v{000}$ and $\v{010}$ [graphs (f) and (h)], as well as $\v{100}$ and $\v{110}$ [graphs (d) and (g)], identical, thus, factoring out the middle qubit. We want the phase factor of $-1$ accumulated for states $\v{000}$ and $\v{010}$ (state $\v{00}$ of the first and the last qubit), and no phase accumulated for states $\v{100}$ and $\v{110}$ (state $\v{10}$ of the first and the last qubit). Graphs (d) and (g) are three-state chain graphs discussed in Sec.~\ref{sec:LG:CH3}, and graphs (f) and (h) are four-state chain graphs investigated in Sec.~\ref{sec:LG:CH4}. Return walks on these graphs require
\begin{eqnarray}\label{eq:2QB:far:system}
\left\{
\begin{array}{l}
|c_1|^2 + |b_3|^2 = k^2,
\\
|c_1|^2 + |b_3|^2 + |a_2|^2 = n^2+m^2,
\\
|c_1|^2|a_2|^2 = n^2m^2,
\\
|m|<|c_1|<|n|,
\\
|m|<|a_2|<|n|,
\end{array}
\right.
\end{eqnarray}
where $n,m$ are integers and $k$ is and even integer. When $n,m$ are odd integers, return walks on graphs (f) and (h) are non-trivial, and a phase of $\pi$ is accumulated. 

As an example we can chose $m=1$, $n=3$, $k=2$, and obtain $|a_2| = \sqrt{6} \approx 2.45$, $|b_3| = \sqrt{5/2} \approx 1.58$, and $|c_1| = \sqrt{3/2} \approx 1.23$. Because all graphs are chain graphs, relative phases are irrelevant and pulse harmonics do not have to be phase locked. The resulting gate is a CZ gate on the first and the last qubits in the three-qubit segment with the first qubit tested for state $\v{0}$ and the $-Z$ gate applied to the last qubit if the test succeeds. Other variations of the CZ gate can be constructed by choosing different transitions.

\subsection{Three-qubit quantum gates}\label{sec:3QB}

Here we investigate an example of a non-trivial entangling three-qubit quantum gate that performs three-qubit Toffoli gate\cite{nielsenchuang} up to two single-qubit Hadamard rotations. A three-qubit Toffoli gate, when represented via CNOT gates, requires at least six CNOT gates applied sequentially.\cite{Markov} A faster implementation that bypasses this limitation is, therefore, beneficial.  Toffoli (or CCNOT) gate can be factored into a sequence 
\begin{eqnarray}\label{eq:3QB:Toffoli}
{\rm Toffoli} = (I\otimes I\otimes{\rm H}){\rm CCZ}(I\otimes I\otimes{\rm H}), 
\end{eqnarray}
where H is the Hadamard gate applied to the third qubit and CCZ is control-Z gate with two control and one target qubits
\begin{eqnarray}\label{eq:3QB:CCZ}
{\rm CCZ} = {\rm diag}(1,1,1,1, 1,1,1,-1).
\end{eqnarray}
As in the case of CZ gates, ``-1'' can be brought to a different location by single-qubit Z gates and the overall phase factor (which is not important in quantum computing). Similarly, the CCZ gates needs {\it return} walks with the adjacency matrix restricted by relation (\ref{eq:2QB:walk}). As an illustration, we will focus on the variation of the CCZ gate in which the amplitude residing on state $\v{100}$ acquires the phase of $\pi$, i.e., 
\begin{eqnarray}
\label{eq:3QB:walk-100}
e^{-i\tau\Lambda}\v{100} &=& -\v{100},\\
\label{eq:3QB:walks-xxx}
e^{-i\tau\Lambda}\v{ijk} &=& \v{ijk},
\quad\quad ijk\neq 100
\end{eqnarray}

\subsection{Completely connected three-qubit system, example}\label{subsec:sym-A}

\begin{figure}\begin{center}
\includegraphics[width=0.9\textwidth]{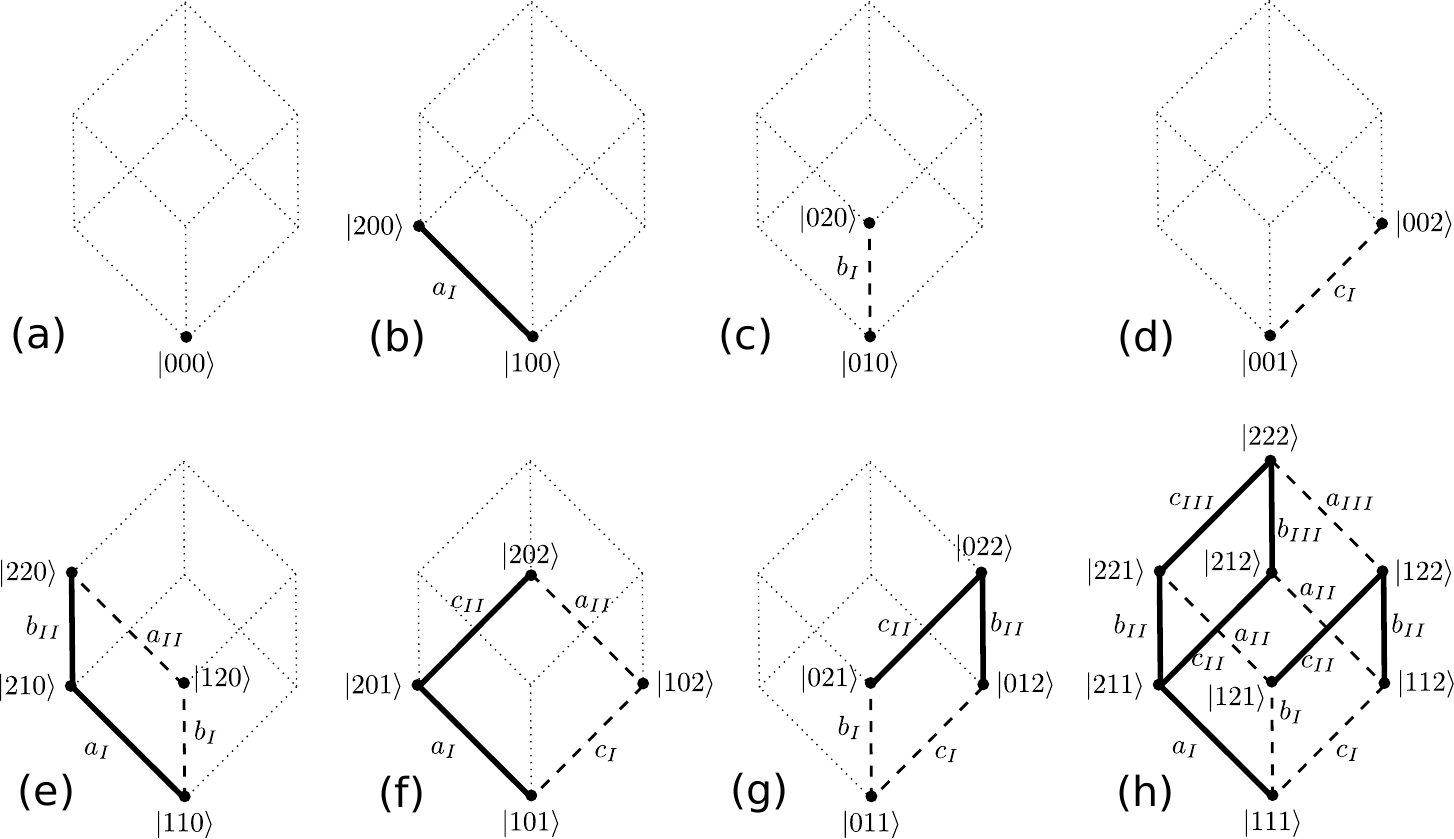}
\vspace*{8pt}
\caption{\label{fig:3QB}
A set of graphs representing a three-qubit system with one active auxiliary state in each qubit and only one allowed transition, $\v{1}\lr\v{2}$, in each qubit system. Dotted lines are guide to the eye. Solid lines indicate resonant transitions activated with external field. Dashed lines are transitions that are allowed but are not used. Dimensionless Rabi frequencies are defined as $\xi = \Omega_\xi\tau/\pi$, where $\xi$ is $a$, $b$, or $c$ with appropriate indexes. Indexes indicate the largest number of auxiliary states in vertexes each transition connects. Note that graphs corresponding to different qubit basis states are not connected with one another because transitions $\v{0}\lr\v{2}$ are not allowed (or not used) in this case.
}
\end{center}\end{figure}

We begin with the symmetry of $\Lambda$ that appears in the case of three qubit systems, e.g., transmons, interacting via a single cavity mode.\cite{solenovNEW} The simplest example of a set of graphs implementing the above evolution in such system is shown in Fig.~\ref{fig:3QB}. We will construct the gate using a single multi-color pulse.
Using a dimensionless representation for each Rabi frequency given by Eq.~(\ref{eq:2QB:Omega_to_a}), as before, we ensure that all walks terminate at the same time $\tau$. We obtain a non-trivial return walk for the graph in Fig.~\ref{fig:3QB}(b) corresponding to Eq.~(\ref{eq:3QB:walk-100}) when
\begin{eqnarray}\label{eq:3QB:n}
|a_I| = n,
\end{eqnarray}
and $n$ is an odd integer (see Sec.~\ref{sec:LG:CH2}). Walks on all other graphs must be trivial return walks. This is trivially the case for graphs (a), (c), (d), and (g), because the corresponding qubit basis states are not connected to any other state by the pulse (corresponding Rabi frequencies are zero). In the case of graphs (e) and (f), a trivial return walk is achieved when
\begin{eqnarray}\label{eq:3QB:m}
\sqrt{|a_I|^2+|b_{II}|^2} = m,
\\\label{eq:3QB:mp}
\sqrt{|a_I|^2+|c_{II}|^2} = m',
\end{eqnarray}
and $m$ and $m'$ are even integers (see Sec.~\ref{sec:LG:CH3}). 

In order to understand the return walk on graph (h), note that it is in fact a square graph (see Sec.~\ref{sec:SQ}) with an additional node attached to it. As explained in Sec.~\ref{sec:SQ}, a square graph can be transformed into a linear chain of four states (see Sec.~\ref{sec:LG:CH4}). Therefore, the entire graph (e) becomes effectively a linear chain of five states discussed in Sec.~\ref{sec:LG:CH5} [see also Fig.~\ref{fig:lin-sq}(d)]. The hopping amplitudes corresponding to this chain are
\begin{eqnarray}
\label{eq:3QB:ampl-a}
\frac{\Omega_a\tau}{\pi} &=& a = a_I,\\
\label{eq:3QB:ampl-b}
\frac{\Omega_b\tau}{\pi} &=& b = 
\sqrt{|b_{II}|^2+|c_{II}|^2}
,\\
\label{eq:3QB:ampl-c}
\frac{\Omega_c\tau}{\pi} &=& c = 
\frac{|b_{II}c_{III}+c_{II}b_{III}|}{|b|}
,\\
\label{eq:3QB:ampl-d}
\frac{\Omega_d\tau}{\pi} &=& d = 
\frac{|c_{II}c_{III}^*-b_{II}b_{III}^*|}{|b|}.
\end{eqnarray}
The walk on such graph returns with trivial phase when
\begin{eqnarray}\label{eq:3QB:kkp}
\left\{
\begin{array}{l}
|a|^2+|b|^2+|c|^2 + |d|^2 = k^2+k'^2
\\
|a|^2|c|^2 + |b|^2|d|^2 + |a|^2|d|^2 = k^2k'^2
\end{array}
\right.
\end{eqnarray}
and $k$ and $k'$ are even integers (see Sec.~\ref{sec:LG:CH5}). This later system of two equation has two unknowns: $|c|$ and $|d|$, and two real parameters: $|a|$ and $|b|$, set by walks on the other graphs. The overall solution for the original Rabi frequencies, however, is not unique because $|c|$ and $|d|$ depend on complex phases of $b_{II}$, $c_{II}$, $b_{III}$, and $c_{III}$ (see Eqs.~\ref{eq:3QB:ampl-c} and \ref{eq:3QB:ampl-d}). In addition, the overall solution is parameterized by five integer parameters $n$, $m$, $m'$, $k$, and $k'$ ($n$ is odd, others are even).

One specific solution mentioned in our earlier work\cite{solenovNEW} that satisfy Eqs.~(\ref{eq:3QB:n})-(\ref{eq:3QB:kkp}) can be obtained assuming all Rabi frequencies are real and $n=1$, $m=m'=2$, $k=2k'=4$. In this case we have
\begin{eqnarray}\nonumber
&&a_I = 1
\\\nonumber
&&b_{II} = c_{II} = \sqrt{3}
\\\label{eq:3QB:solution0}
&&b_{III} = \frac{3+\sqrt{17}}{2}
\\\nonumber
&&c_{III} = \frac{3-\sqrt{17}}{2}
\end{eqnarray}

\begin{figure}\begin{center}
\includegraphics[width=0.9\textwidth]{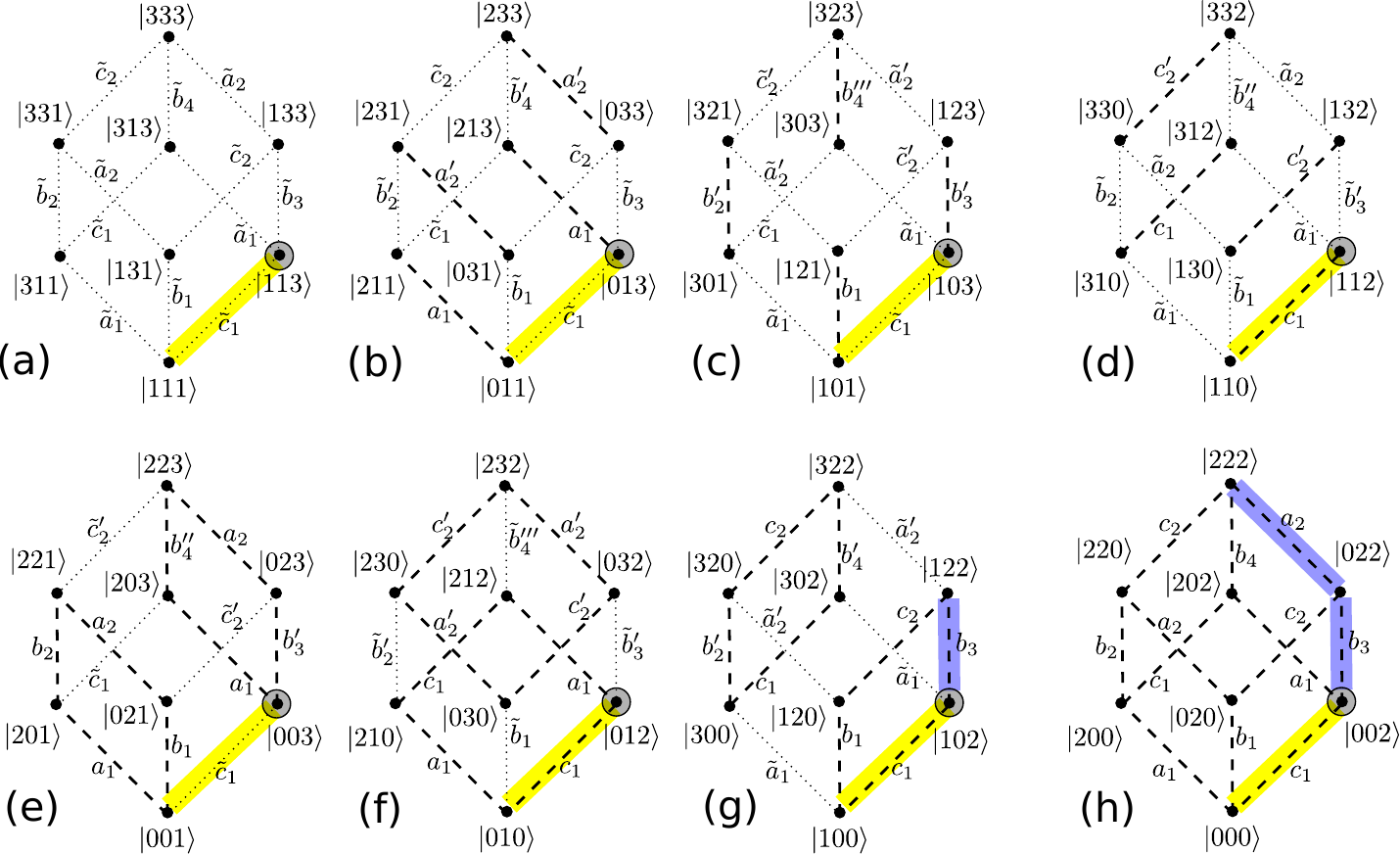}
\vspace*{8pt}
\caption{\label{fig:3QB-QDs}
A set of graphs based on Fig.~\ref{fig:QDs-3QB}(j) representing a three-qubit subsystem of the quantum dot register with two allowed transitions $\v{0}\lr\v{2}$ and $\v{1}\lr\v{3}$ in each dot. The sub-graphs cut by the choice of the pulse harmonics are highlighted (in yellow for the first and the last pulse; in blue for the middle pulse). The symmetry of the transition amplitudes (Rabi frequencies) is shown via dimensionless amplitudes defined by Eq.~(\ref{eq:2QB:Omega_to_a}). 
}
\end{center}\end{figure}

\subsection{Scalable system, three-qubit subset}\label{subsec:sym-B}

As mentioned in the previous section, single-cavity systems are not scalable and can not accommodate large number of qubits due to spectral crowding. Scalable  multi-cavity registers discussed in the previous section induce symmetry reduction that is different form the one used in Sec.~\ref{subsec:sym-A} above. As an example, we will consider a network produce by a three-qubit segment of the quantum dot register shown in Figs.~\ref{fig:QDs-3QB}(i-j). When only transitions originating from single qubit $\v{0}\lr\v{2}$ and $\v{1}\lr\v{3}$ transitions are involved, the three-qubit subgraph splits into a set of cube graphs shown in Fig.~\ref{fig:3QB-QDs}. The symmetry of the system is such that the connections opposite to each other on the top as well as the bottom faces of each cube are indistinguishable in the intermediate resonance regime. In addition, connections involving nodes that differ only by one index, i.e., $\v{i02}\lr\v{i22}$ for various $i$, are also indistinguishable, except for those with excitations in all three qubit systems (see Sec.~\ref{sec:Systems:QDs:three} for derivation). The symmetry is displayed in Fig.~\ref{fig:3QB-QDs} using reduced Rabi frequency labels defined by Eq.~(\ref{eq:2QB:Omega_to_a}). 

With this reduction of symmetry (\ref{eq:QB:NI-Omega}), it is most natural to perform various CCZ gates using three pulses with two frequencies each. Examples of pulse compositions that produce different variations of the CCZ gate are given in Table~\ref{tab:CCZ}. Here we give a detailed discussion of the fifth example, when pulses accumulate a non-trivial phase for basis state $\v{100}$ as before. The first and the last pulses are identical and activate only one connection for each cube graph in Fig.~\ref{fig:3QB-QDs}. They are simple $\pi$-pulses performed concurrently, i.e., we set $c_1=\tilde c_1 = 1/2$. After the first two-color pulse, the population is moved to the bottom right (highlighted) node in each cube. The population is returned back to the qubit states by the last two-color pulse. This means that all quantum walks induced by the middle two-color pulse must start from and return to those highlighted nodes. By setting $b_3 = 1$ we produce a simple non-trivial return walk for node $\v{102}$ in graph (g). In order to produce a trivial walk for graph (h), which also has $b_3$ connection activated, we set $a_2=\sqrt{3}$. This reduces graph (h) to a chain of three states discussed in Sec.~\ref{sec:LG:CH3} below. Note that all other connections remain inactive because they are off resonance with narrow-bandwidth pulse harmonics used for $b_3$ and $a_2$. This leads to trivial evolution on graphs (a-f) during the second two-color pulse. As the result, the amplitude residing on state $\v{100}$ accumulates a phase of $2\pi$ (a factor of $+1$) and all other qubit states accumulate a phase of $\pi$ (a factor of $-1$). Because the overall phase factor is negligible, this is equivalent to a CCZ gate with control triggered by $0$ in the second and the third qubit and Z applied to the first qubit. Other CCZ gates are obtained similarly, activating different transition in the same fashion. Note that in all these examples evolution does not depend on phases of the entries of the adjacency matrix (Rabi frequencies), and, hence, all three two-color pulses do not have to be phase-locked. Generally this is not the case if loop graphs, e.g., shown in Fig.~\ref{fig:lin-sq}(f), are involved.

\begin{table}
\centering\scriptsize
\begin{tabular}{|l|l|l|l|}
\hline
{\bf CCZ gate diagonal$^*$}&
{\bf two-color pulse sequence; not phase-locked}&
{\bf gate type}&
{\bf state}
\\\hline
$\{-1,1,1,1,1,1,1,1\}$ &
$\{\{c_1 = \tilde c_1 = 1/2 \}_1,
\{\tilde a'_2 = \sqrt{3}; b_3 = 1 \}_2,
\{c_1 = \tilde c_1 = 1/2 \}_3\}$&
$\bar{\rm C} \bar{\rm C} \bar{\rm Z}$&
$\v{000}$
\\
$\{1,-1,1,1,1,1,1,1\}$ &
$\{\{c_1 = \tilde c_1 = 1/2 \}_1,
\{\tilde a'_2 = \sqrt{3}; b'_3 = 1 \}_2,
\{c_1 = \tilde c_1 = 1/2 \}_3\}$&
$\bar {\rm C} \bar {\rm C} {\rm Z}$&
$\v{001}$
\\
$\{1,1,-1,1,1,1,1,1\}$ &
$\{\{c_1 = \tilde c_1 = 1/2 \}_1,
\{\tilde a_2 = \sqrt{3}; \tilde b'_3 = 1 \}_2,
\{c_1 = \tilde c_1 = 1/2 \}_3\}$&
$\bar{\rm C} {\rm Z} \bar{\rm C}$&
$\v{010}$
\\
$\{1,1,1,-1,1,1,1,1\}$ &
$\{\{c_1 = \tilde c_1 = 1/2 \}_1,
\{\tilde a_2 = \sqrt{3}; \tilde b_3 = 1 \}_2,
\{c_1 = \tilde c_1 = 1/2 \}_3\}$&
$\bar{\rm Z} {\rm C} {\rm C}$&
$\v{011}$
\\
$\{1,1,1,1,-1,1,1,1\}$ &
$\{\{c_1 = \tilde c_1 = 1/2 \}_1,
\{a_2 = \sqrt{3}; b_3 = 1 \}_2,
\{c_1 = \tilde c_1 = 1/2 \}_3\}$&
${\rm Z} \bar{\rm C} \bar{\rm C}$&
$\v{100}$
\\
$\{1,1,1,1,1,-1,1,1\}$ &
$\{\{c_1 = \tilde c_1 = 1/2 \}_1,
\{a_2 = \sqrt{3}; b'_3 = 1 \}_2,
\{c_1 = \tilde c_1 = 1/2 \}_3\}$&
${\rm C} \bar{\rm Z} \bar{\rm C}$&
$\v{101}$
\\
$\{1,1,1,1,1,1,-1,1\}$ &
$\{\{c_1 = \tilde c_1 = 1/2 \}_1,
\{a'_2 = \sqrt{3}; \tilde b'_3 = 1 \}_2,
\{c_1 = \tilde c_1 = 1/2 \}_3\}$&
${\rm C} {\rm C} \bar{\rm Z}$&
$\v{110}$
\\
$\{1,1,1,1,1,1,1,-1\}$ &
$\{\{c_1 = \tilde c_1 = 1/2 \}_1,
\{a'_2 = \sqrt{3}; \tilde b_3 = 1 \}_2,
\{c_1 = \tilde c_1 = 1/2 \}_3\}$&
${\rm C} {\rm C} {\rm Z}$&
$\v{111}$
\\\hline 
\end{tabular}
{\\$^*$ the basis is $\{000,001,010,011,100,101,110,111\}$.}
\caption{Examples of three-qubit CCZ gates (diagonal Toffoli gates) performed via three pulses of two frequencies each. A subscript to a group indicates pulse order in temporal sequence. Non-zero adjacency matrix entries (Rabi frequencies) are shown for each pulse. Dimensional Rabi frequencies can be obtained using Eq.~(\ref{eq:2QB:Omega_to_a}).  All pulses are resonant and do not have to be phase-locked. Note that in all given examples, pulse harmonics are applied to each dot ($a,b,c$).  Other quantum-walk-based sequences performing the same gates, e.g., based on $a_1$ and $\tilde a_1$, can be constructed in a similar fashion. Some may require pulses to be phase locked if complex phases of the adjacency matrix entries matter. The table also shows the basis state for which the phase of $\pi$ is accumulated, and control-target qubit order, where the over-bar denotes application of control (C) or sign change (Z) to $\v{0}$ instead of $\v{1}$. 
}
\label{tab:CCZ}
\end{table}

\subsection{Gate performance}
\label{sec:time-comparison}

It is instructive to investigate the execution time of a three-qubit gate performed using quantum walks, for example the CCZ gate constructed in Sec.~\ref{subsec:sym-B}. The best known decomposition of a three-qubit Toffoli gate using CNOT gates involves at least six CNOT gates. Thus, if performed via a CZ gates, each CCZ gate requires at least six CZ gates applied to different pairs of qubits (two for each pair). This provides a reference point for the execution time of similar quantum-walk-based gate.

In scalable systems outlined in Sec.~\ref{sec:Systems}, two-qubit CZ gates between adjacent qubits are performed via four $\pi$ pulses, as described in Refs.~\onlinecite{solenov1, solenov2,solenov3}, or a single multi-color pulse, as described in Sec.~\ref{sec:2QB}. Following Sec.~\ref{sec:QW}, we can use duration and control field amplitude of the $\pi$ pulses as 1/2 of a unit of time and a unit of field amplitude respectively, and rescale the amplitudes of the multi-color pulses such that amplitude of each Fourier harmonic does not exceed those of the $\pi$ pulses. In this case the relative execution time of the CZ gate performed via four $\pi$ pulses is $t_g/2t_\pi = 2$. The largest amplitude in the first example in Sec.~\ref{sec:2QB} is $2\pi/\tau$ which corresponds to $t_g/2t_\pi=2$ (also $\approx 2.7\pi/\tau$ and $t_g/2t_\pi=2.7$ for another example in the same section). The CZ gates applied to the next nearest neighbors along the chain register are constructed in Sec.~\ref{sec:2QB-far} and have the largest Rabi frequency $\sqrt{6}\pi/\tau$ which corresponds to $t_g/2t_\pi\approx 2.45$.

Three-qubit gates described in Sec.~\ref{subsec:sym-B} use three two-color pulses to perform quantum walks. The first and the last pulses are equivalent to pairs of concurrent $\pi$ pulses. The middle pulse has the largest amplitude $\sqrt{3}\pi/\tau$, which results in the duration of $\approx 1.7\times 2t_\pi$ after rescaling. The total duration of the gate is $t_g/2t_\pi \approx 1/2 + 1.7 + 1/2 = 2.7$.  It follows that the execution time is longer then that of a CZ gate but faster than that of two CZ gates, i.e., $t_g({\rm CZ})<t_g({\rm CCZ})<2t_g({\rm CZ})$. The standard decomposition into the CZ gates requires $t_g({\rm CCZ})\ge 6t_g({\rm CZ})$. Therefore CCZ and Toffoly gates performed using quantum walks can be executed more than three times faster as compared to standard CNOT decomposition in the same system given the maximum allowed control field amplitude for each pulse harmonic. Note that the speed of the CCZ gate performed using quantum walks depends on the symmetry of the graphs (degree of physical interaction in the system), the choice of sub-graphs for each walk, the number of (multi-color) pulses, as well as other parameters. In particular, complex single-pulse walks may, in some cases, require large adjacency matrix entries, which will lead to longer gate time (rescaling the amplitudes).

\section{Linear graphs}\label{sec:LG}

In this section we discuss return quantum walks on graphs that are a set of states connected in linear chains of various (finite) length. These graphs can be used to implement single-qubit gates (see Sec.~\ref{sec:1QB}) and are essential building blocks for more complex graphs needed to implement entangling quantum gates (see Secs.~\ref{sec:2QB} and \ref{sec:3QB}).

Two distinct types of return walks in these systems will be emphasized: (i) trivial return walk, ${\cal R}_0$, in which amplitudes returns back to initial state and acquire no phase, and (ii) non-trivial return walk, ${\cal R}_\pi$, in which phase $\pi$ is accumulated when the system returns back to its initial state. The second type, ${\cal R}_\pi$ walk, is only possible if adjacency matrix is not singular, which can be readily verified by calculating the evolution operator via eigendecomposition (see examples below). In general, it is, therefore, expected that linear chain graphs with even number of vertexes can support both ${\cal R}_0$ and ${\cal R}_\pi$ walks (i.e., both $U=\pm 1$ are possible), while linear chain graphs with odd number of vertexes can not produce $U=-1$. The latter statement, does not strictly eliminate ${\cal R}_\pi$ walks, as a possibility that ${\cal R}_\pi$ exists, even when $U\neq-1$, remains, in principle.

Four examples of linear chain graphs are discussed below. Detailed derivation of the walks are given in Appendixes~\ref{app:LG:CH3}-\ref{app:LG:CH5}. Some more complex graphs, which appear in two- and three-qubit gates and simplify to linear chain graphs, are discussed in the next sections.  

\subsection{chain of two states}
\label{sec:LG:CH2}

The simplest linear chain graph that we discuss here briefly for completeness is the one that corresponds to a two-state quantum system [see Fig.~\ref{fig:lin-sq}(a)]. The adjacency matrix is
\begin{eqnarray}\label{eq:LG:CH2:Lambda}
\Lambda =
\left(\begin{array}{cc}
0 & \Omega_a \\
\Omega_a^* & 0
\end{array}\right)
\equiv\frac{\pi}{\tau}
\left(\begin{array}{cc}
0 & a \\
a^* & 0
\end{array}\right)
\end{eqnarray}
The evolution operator can be easily found by direct re-summation of odd and even terms of the exponential series
\begin{eqnarray}\label{eq:LG:CH2:U}
U(\tau) = e^{-i\tau\Lambda}=
\cos\pi|a|-i\frac{\Lambda}{|\Omega_a|}\sin\pi|a|
\end{eqnarray}
The ``return'' condition, which is identical to condition (\ref{eq:QW:QLP}) in this case, is satisfied provided $|a|$ is an integer number. We have
\begin{align}\label{eq:LG:CH2:R_0}
\text{\bf Walk {${\cal R}_0$}:}\;\;\;&
|\Omega_a| \tau/\pi = |a| = 2n,
\quad & n\in \mathbb{Z}
\\\label{eq:LG:CH2:R_pi}
\text{\bf Walk {${\cal R}_\pi$}:}\;\;\;&
|\Omega_a| \tau/\pi = |a| = 2n+1,
\quad & n\in \mathbb{Z}
\end{align}
where the former defines a trivial return walk and the latter defines a non-trivial return walk that accumulates the phase of $\pi$.

\subsection{chain of three states}
\label{sec:LG:CH3}

\begin{figure}\begin{center}
\includegraphics[width=0.7\columnwidth]{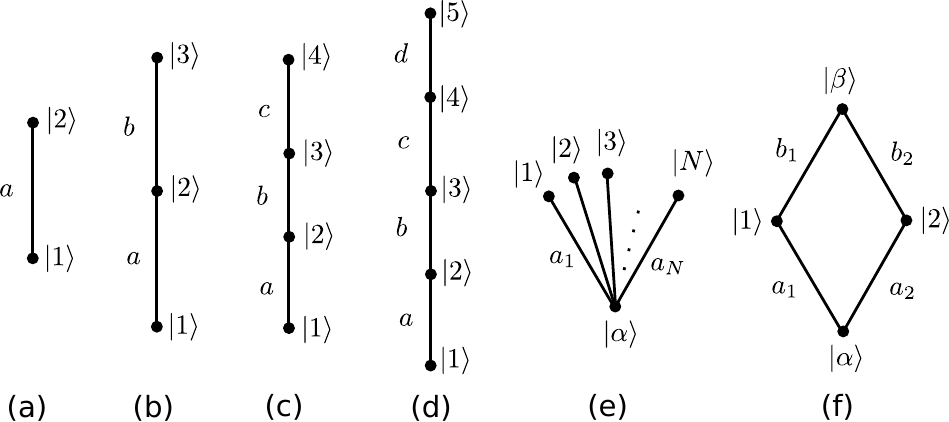}
\caption{\label{fig:lin-sq}
Linear chain graphs (a-d), a fan/tree graph (e), and a square graph (f); for details see Secs.~\ref{sec:LG:CH2}~-~\ref{sec:LG:CH5}, Sec.~\ref{sec:TG:F}, and Sec.~\ref{sec:SQ}, respectively. In each case, dimensionless amplitudes are defined as $\xi = \Omega_\xi \tau / \pi$, where $\Omega_\xi$ are Rabi frequencies (due to external control) corresponding to each transition, and $\xi$ stand for $a$, $b$, $c$, etc.
}\end{center}\end{figure}

A graph of three states connected in a chain [see Fig.~\ref{fig:lin-sq}(b)] is described by the adjacency matrix
\begin{eqnarray}\label{eq:LG:CH3:Lambda}
\Lambda =
\left(\begin{array}{ccc}
0 & \Omega_a & 0 \\
\Omega_a^* & 0 & \Omega_b \\
0 & \Omega_b^* & 0
\end{array}\right)
\equiv\frac{\pi}{\tau}
\left(\begin{array}{ccc}
0 & a & 0 \\
a^* & 0 & b \\
0 & b^* & 0
\end{array}\right)
\end{eqnarray}
The evolution operator corresponding to this system can be found explicitly as in the previous case, although the exact expression becomes cumbersome, see Appendix~\ref{app:LG:CH3}. A more elegant approach, also applicable to larger systems, is to notice that diagonalization of (\ref{eq:LG:CH3:Lambda})
\begin{eqnarray}\label{eq:LG:CH3:eig}
M^\dag\Lambda M = {\rm diag}(\lambda_i)
\end{eqnarray}
yields three eigenvalues 
\begin{eqnarray}\label{eq:LG:CH3:lambda}
\lambda_i = \{
-\sqrt{|\Omega_a|^2+|\Omega_b|^2},
0,
+\sqrt{|\Omega_a|^2+|\Omega_b|^2},
\}
\end{eqnarray}
one of which is zero. The evolution operator, therefore, becomes
\begin{eqnarray}\label{eq:LG:CH3:U}
U = M
\left(\begin{array}{ccc}
e^{-i\pi\sqrt{|a|^2+|b|^2}} & 0 & 0 \\
0 & 1 & 0 \\
0 & 0 & e^{+i\pi\sqrt{|a|^2+|b|^2}}
\end{array}\right)
M^\dag
\end{eqnarray}
As the result, irrespective of $M$, we have
\begin{align}\label{eq:LG:CH3:R_0}
\text{\bf Walk {${\cal R}_0$}:}\;\;\;&
U=1\to \sqrt{|a|^2+|b|^2} = 2n,
\, & n\in \mathbb{Z}
\\\label{eq:LG:CH3:R_pi}
\text{\bf Walk {${\cal R}_\pi$}:}\;\;\;&
U=-1\to\text{not achievable}
\quad &
\end{align}
This, strictly speaking, does not imply that ${\cal R}_\pi$ is not possible. However by inspecting the complete solution (see Appendix~\ref{app:LG:CH3}) we see that ${\cal R}_\pi$ walk is not accessible in this system if the initial state is state $\v{1}$ or $\v{3}$.

\vspace*{12pt}
\noindent
{\bf Special case: integer amplitudes}. It is interesting to note that the system allows integer amplitudes, which can become useful when implementing quantum gates via quantum walks on graphs with small number of free parameters. Note that $|a|^2+|b|^2 = n^2$ describes Pythagorean triples (or triangles), therefore
\begin{eqnarray}\label{eq:LG:CH3:tri}
\left\{\begin{array}{l}
|a| = i^2-j^2 \\
|b| = 2ij \\
n = i^2 + j^2
\end{array}\right.
\quad ij \in \mathbb{Z}>0
\end{eqnarray}
Irreducible triples have odd $n$ and, therefore, must be multiplied by an even integer to produce ${\cal R}_0$ walk (\ref{eq:LG:CH3:R_0}), e.g., 
$\{|a|,|b|,n\}\to 2\{3,4,5\}$, $\{|a|,|b|,n\}\to 2\{5,12,13\}$, $\{|a|,|b|,n\}\to 2\{8,15,17\}$.

\subsection{chain of four states}\label{sec:LG:CH4}

The adjacency matrix of a chain of four states [see Fig.~\ref{fig:lin-sq}(c)] is
\begin{eqnarray}\label{eq:LG:CH4:Lambda}
\Lambda =
\left(\begin{array}{cccc}
0 & \Omega_a & 0 & 0 \\
\Omega_a^* & 0 & \Omega_b & 0 \\
0 & \Omega_b^* & 0 & \Omega_c \\
0 & 0 & \Omega_c^* & 0
\end{array}\right)
\equiv\frac{\pi}{\tau}
\left(\begin{array}{cccc}
0 & a & 0 & 0 \\
a^* & 0 & b & 0 \\
0 & b^* & 0 & c \\
0 & 0 & c^* & 0
\end{array}\right)
\end{eqnarray}
Due to symmetry, the eigenvalues are $\pm \lambda_1$ and $\pm \lambda_2$. The system does not have zero eigenvalues and, thus, unlike in the previous case shown in Eq.~(\ref{eq:LG:CH3:U}), does allow $U=1$ and $U=-1$ evolutions. By setting $\lambda_1\tau=\pi n$ and $\lambda_2\tau=\pi m$, where $n$ and $m$ are integers of the same parity, we obtain (see Appendix~\ref{app:LG:CH4})
\begin{eqnarray}\label{eq:LG:CH4:system}
\left\{\begin{array}{l}
|a|^2+|b|^2+|c|^2 = n^2+m^2 \\
|a||c| = |n||m|
\end{array}\right.
\quad n,m \in \mathbb{Z}
\end{eqnarray}
As the result, we have
\begin{align}\label{eq:LG:CH4:R_0}
\text{\bf Walk {${\cal R}_0$}:}\;\;\;
U=1,
\quad&\text{Eq.~(\ref{eq:LG:CH4:system}) with $n,m \in $ even}&
\\\label{eq:LG:CH4:R_pi}
\text{\bf Walk {${\cal R}_\pi$}:}\;\;\;
U=-1,
\,&\text{Eq.~(\ref{eq:LG:CH4:system}) with $n,m \in$ odd}&
\end{align}

It is interesting to note that if we solve system (\ref{eq:LG:CH4:system}) for $b$ we obtain
\begin{eqnarray}\label{eq:LG:CH4:b}
|b|=\sqrt{
(|n|+|m|)^2 - \left(\frac{|nm|}{|a|}+|a|\right)^2
}
\end{eqnarray}
This defines the range of valid values for $|a|$ and $|c|$
\begin{eqnarray}\label{eq:LG:CH4:a_range}
|m| \le |a| \le |n|,\quad
|m| \le |c| \le |n|,
\quad\quad |n|>|m|.
\end{eqnarray}
The system becomes disconnected into a pair of two-state systems, when $|n|=|m|$. 

\vspace*{12pt}
\noindent
{\bf Special case: integer amplitudes}. Integer amplitudes can be used in the symmetric case when $|a|^2=|c|^2$. From system (\ref{eq:LG:CH4:system}) we obtain
\begin{eqnarray}\label{eq:LG:CH4:int-sys}
\left\{\begin{array}{l}
|a|^2 = |n||m| \\
|b| = ||n|-|m||
\end{array}\right.
\end{eqnarray}
Therefore, for any positive integer $|a|$, one can define integer amplitudes
\begin{eqnarray}\label{eq:LG:CH4:int}
\left\{\begin{array}{l}
|a|\in \mathbb{Z}\\
|b| = ||a|^2-1| \\
|c| = |a|
\end{array}\right.
\end{eqnarray}

\subsection{chain of five states}\label{sec:LG:CH5}

The adjacency matrix of a chain of five states shown in Fig.~\ref{fig:lin-sq}(d)
\begin{eqnarray}\label{eq:LG:CH5:Lambda}
\left(\begin{array}{ccccc}
0 & \Omega_a & 0 & 0 & 0 \\
\Omega_a^* & 0 & \Omega_b & 0 & 0 \\
0 & \Omega_b^* & 0 & \Omega_c & 0 \\
0 & 0 & \Omega_c^* & 0 & \Omega_d \\
0 & 0 & 0 & \Omega_d^* & 0
\end{array}\right)
\equiv\frac{\pi}{\tau}
\left(\begin{array}{ccccc}
0 & a & 0 & 0 & 0\\
a^* & 0 & b & 0 & 0\\
0 & b^* & 0 & c & 0\\
0 & 0 & c^* & 0 & d\\
0 & 0 & 0 & d^* & 0
\end{array}\right)
\end{eqnarray}
is singular. One of the eigenvalues is zero, and the remaining eigenvalues are $\pm \lambda_1$ and $\pm \lambda_2$, where $\lambda_{1,2}$ are given in Appendix~\ref{app:LG:CH5}. The evolution operator is a trivial identity matrix if $\lambda_1\tau=\pi n$ and $\lambda_2\tau=\pi m$, where $n$ and $m$ are even integers. Solving these two equation we obtain
(see Appendix~\ref{app:LG:CH5}) 
\begin{eqnarray}\label{eq:LG:CH5:system}
\left\{\begin{array}{l}
|a|^2+|b|^2+|c|^2 + |d|^2 = n^2+m^2 \\
|a|^2|c|^2 + |b|^2|d|^2 + |a|^2|d|^2 = n^2m^2
\end{array}\right.
\quad n,m \in \mathbb{Z}
\end{eqnarray}
As the result, we have
\begin{align}\label{eq:LG:CH5:R_0}
\text{\bf Walk {${\cal R}_0$}:}\;\;\;&
U=1 \to \text{Eq.~(\ref{eq:LG:CH5:system}) with $n,m \in $ even}
\\\label{eq:LG:CH5:R_pi}
\text{\bf Walk {${\cal R}_\pi$}:}\;\;\;&
U=-1 \to \text{not achievable}
\end{align}

\vspace*{12pt}
\noindent
{\bf Special case: integer amplitudes}. Integer amplitudes are possible for a symmetric chain with $|a|=|d|$ and $|b|=|c|$. We obtain
\begin{eqnarray}\label{eq:LG:CH5:int-sys}
\left\{\begin{array}{l}
|a| = |m| \\
|b|^2 = (n^2-m^2)/2
\end{array}\right.
\quad\quad n>m\in {\rm even}
\end{eqnarray}

\section{Single level tree (fan) graphs}\label{sec:TG}
\label{sec:TG:F}

These graphs are structures that branch from a single vertex, as shown in Fig.~\ref{fig:lin-sq}(e). They are natural parts of a network of transitions in multiqubit systems with at least one ``local'' transition allowed for each qubit. The corresponding adjacency matrix is
\begin{eqnarray}\label{eq:TG:F:Lambda}
\Lambda = \sum_{j=1}^N \Omega_j\v{\alpha}\iv{j}+ h.c.
= \frac{\pi}{\tau}\sum_{j=1}^N a_j\v{\alpha}\iv{j}+ h.c.
\end{eqnarray}
Quantum evolution on such graphs can be mapped onto that of a two state quantum system.
To perform the map, define state
\begin{eqnarray}\label{eq:TG:F:s}
a^*\v{s} = \sum_{j=1}^N a^*_j\v{j}
\end{eqnarray}
such that $\av{s|s}=1$, i.e., 
\begin{eqnarray}\label{eq:TG:F:x2}
|a|^2 = \sum_{j=1}^N |a_j|^2
\end{eqnarray}
As the result the adjacency matrix becomes
\begin{eqnarray}\label{eq:TG:F:Lambda}
\Lambda = \frac{\pi}{\tau} a \v{\alpha}\iv{s}+ h.c.
\end{eqnarray}
which is identical to (\ref{eq:LG:CH2:Lambda}). We obtain
\begin{align}\label{eq:TG:F:R_0}
\text{\bf Walk {${\cal R}_0$}:}\;\;\;&
\sqrt{\sum_{j=1}^N |a_j|^2} = 2n,
\quad & n\in \mathbb{Z}
\\\label{eq:TG:F:R_pi}
\text{\bf Walk {${\cal R}_\pi$}:}\;\;\;&
\sqrt{\sum_{j=1}^N |a_j|^2} = 2n+1,
\quad & n\in \mathbb{Z}
\end{align}

\section{A square graph}\label{sec:SQ}

This is the graph [see Fig.~\ref{fig:lin-sq}(f)] that naturally appears in the simplest quantum walk implementation of a CZ gate in the system with one (active) auxiliary state per qubit. The adjacency matrix is
\begin{eqnarray}\nonumber
\Lambda &=& 
\Omega_1\v{\alpha}\iv{1} \!+\! \Omega_2\v{\alpha}\iv{2}
\!+\! 
\Omega'_1\v{1}\iv{\beta} \!+\! \Omega'_2\v{2}\iv{\beta}
\!+\! h.c.
\\\label{eq:SQ:Lambda}
&=&
\frac{\pi}{\tau}\!\left(
\phantom{\frac{0}{0}}\!\!\!\!\!
a_1\v{\alpha}\iv{1} \!+\! a_2\v{\alpha}\iv{2}
\!+\!
b_1\v{1}\iv{\beta} \!+\! b_2\v{2}\iv{\beta}
\!+\! h.c.
\right)
\end{eqnarray}
Similarly to symmetric fan graphs discussed above, it can be mapped onto a graph representing linear chain of states. We can define state $\v{s}$ such that
\begin{eqnarray}\label{eq:SQ:s}
s^*\v{s} = a^*_1\v{1}+a^*_2\v{2},
\quad\quad
|s|^2 = |a_1|^2 + |a_2|^2
\end{eqnarray}
In this case, the adjacency matrix transforms to
\begin{eqnarray}\label{eq:SQ:Lambda-t}
\frac{\Lambda\tau}{\pi}=
s\v{\alpha}\iv{s}
\!+\!
b_1\frac{a_1s^*\v{s}+a_2^*s\v{a}}{|s|^2}\iv{\beta}
+ b_2\frac{a_2s^*\v{s}-a_1^*s\v{a}}{|s|^2}\iv{\beta}
\!+\! h.c.
\end{eqnarray}
where state
\begin{eqnarray}\label{eq:SQ:a}
\v{a} = (a_2\v{1}-a_1\v{2})/s
\end{eqnarray}
is orthogonal to $\v{s}$. This gives to possibilities.

\subsection{symmetric case}\label{sec:SQ:S}

One possibility is to have $b_1a_2^* = b_2a_1^*$, in which case
\begin{eqnarray}\label{eq:SQ:S:Lambda}
\frac{\Lambda\tau}{\pi}=
s\v{\alpha}\iv{s}
\!+\!
s'\v{s}\iv{\beta}
\!+\! h.c.,
\end{eqnarray}
where
\begin{eqnarray}\label{eq:SQ:S:s'}
s'=s^*\frac{a_1b_1+a_2b_2}{|s|^2}
\end{eqnarray}
The adjacency matrix becomes that of the linear chain graph of three states. In this case the solution is
\begin{align}\label{eq:SQ:S:R_0}
\text{\bf Walk {${\cal R}_0$}:}\;\;\;&
U=1\to \sqrt{|s|^2+|s'|^2} = 2n,
& n\in \mathbb{Z}
\\\label{eq:SQ:S:R_pi}
\text{\bf Walk {${\cal R}_\pi$}:}\;\;\;&
U=-1\to\text{not achievable}
\quad &
\end{align}
as obtained earlier in Sec.~\ref{sec:LG:CH3}.

\subsection{non-symmetric case}\label{sec:SQ:N}

The other possibility is to have $b_1a_2^* \neq b_2a_1^*$. In this case, adjacency matrix is
\begin{eqnarray}\label{eq:SQ:N:Lambda}
\frac{\Lambda\tau}{\pi}=
s\v{\alpha}\iv{s}
\!+\!
s'\v{s}\iv{\beta}
\!+\!
a\v{\beta}\iv{a}
\!+\! h.c.
\end{eqnarray}
where
\begin{eqnarray}\label{eq:SQ:N:a}
a=s^*\frac{a_2b_1^*-a_1b_2^*}{|s|^2}
\end{eqnarray}
This corresponds to the linear chain graph of four states discussed in Sec.~\ref{sec:LG:CH4}. The solution is
\begin{eqnarray}\label{eq:SQ:N:system}
\left\{\begin{array}{l}
|s|^2+|s'|^2+|a|^2 = n^2+m^2 \\
|s||a| = |n||m|
\end{array}\right.
\quad n,m \in \mathbb{Z}
\end{eqnarray}
and
\begin{align}\label{eq:SQ:N:R_0}
\text{\bf Walk {${\cal R}_0$}:}\;\;\;&
U=1,&\text{Eq.~(\ref{eq:SQ:N:system}) with $n,m \in $ even}
\\\label{eq:SQ:N:R_pi}
\text{\bf Walk {${\cal R}_\pi$}:}\;\;\;&
U=-1,&\text{Eq.~(\ref{eq:SQ:N:system}) with $n,m \in$ odd}
\end{align}
as obtained earlier in Sec.~\ref{sec:LG:CH4}. In this case a non-trivial return walk is possible.

\subsection{partitioning}\label{sec:SQ:P}

It is interesting to note that the square graph can be partitioned into two non-trivial two-node graphs, one having state $\v{\alpha}$ and the other having state $\v{\beta}$, in an infinite number of ways. This is achieved by setting hopping amplitude $s'$ to zero, or, explicitly, by requiring
\begin{eqnarray}\label{eq:SQ:P:condition}
a_1b_1+a_2b_2 = 0.
\end{eqnarray}
The two specific straightforward cases of partitioning are obtained for $a_1=b_2=0$ or $a_2=b_1=0$.

Such partitioning, as also discussed in Refs.~\refcite{Cavin,Loke}, can be particularly advantageous when performing similar analysis for larger graphs, for which analytical solutions may be difficult or impossible to find. This approach can simplify construction of multiqubit gates performed via quantum walks in larger systems.

\section{Conclusion}\label{sec:conclude}

We introduced a new scalable approach to quantum gates based on continuous time quantum walks. This approach relies on availability of auxiliary, typically higher energy, states that can take part in interactions between qubit systems. Interacting excited states create a developed network of states, see, e.g., Fig.~\ref{fig:QDs-2QB}(g) and \ref{fig:QDs-3QB}(j), through which entanglement can propagate. This additional resource---an interacting entangling bus---potentially enables much faster entanglement propagation and  quantum gates. As an example, Toffoli gate that needs a minimum of six CNOT gates\cite{nielsenchuang} to be implemented can run as fast as $\sim 1.35$ of a run-time of a single CNOT gate, as analyzed in Sec.~\ref{sec:time-comparison}. Such dramatic compression of multiqubit gates does not rely solely on the existence of entanglement bus, but require an efficient way to probe physical non-local interactions present there.

Many systems do have multiple well-defined excited states that can mediate interactions between qubits. However,
as demonstrated in Sec.~\ref{sec:Systems} and subsections therein, interactions in a scalable qubit register are restricted by certain symmetry. The symmetry originates from that of a non-interacting collection of multi-state quantum systems, see Eqs.~(\ref{eq:QB:NI-Omega}), and is subsequently reduced when physical interactions are present, but not lifted entirely. As the result, entanglement bus becomes a complex collection of states each span over few qubit systems (due to physical interactions), e.g., as shown in Fig.~\ref{fig:QDs-3QB}(j). Traditional approach---driving system through a well defined trajectory---is possible but does not give any gate compression.\cite{solenovNEW} Compression becomes possible when multiple trajectories are addressed concurrently, simultaneously probing interactions present in different parts of the spectrum. This is accomplished using {\it continuous time quantum walks}, see Sec.~\ref{sec:QW}.

Unlike in many other proposed implementations of {\it continuous time quantum walks}, here the walk does not propagate in real time. Evolution takes place in an effective time, see Eq.~(\ref{eq:QW:walk}), generated by external control pulses, see Eqs.~(\ref{eq:QW:V-gen}), (\ref{eq:QW:tau}), (\ref{eq:QW:V-multi}), and (\ref{eq:QW:tau-multi}). This effective time can {\it begin} and {\it end}, enabling walks of precise duration. Solution to a coherent quantum walk, given the network of states (graph) with corresponding hopping amplitudes, is rather easy to find. The problem of constructing an entangling gate, however, is of different type. A quantum gate is defined via a set of boundary conditions in the (effective) time---the initial and the final conditions, see, e.g., Eqs.~(\ref{eq:2QB:walk-00}). These conditions must be satisfied to perform the desired gate. In addition, there are symmetry relations on hopping amplitudes enforced by physical interactions and configuration (connectivity) of the entanglement bus. The problem therefore is to find quantum walk solutions that make these sets of restrictions consistent. This is not always possible: as an example, the set of symmetry conditions (\ref{eq:QB:NI-Omega}) corresponding to non-interacting qubit systems is inconsistent with any time-domain boundary conditions required for entangling quantum gates. Further complexity comes from the fact that a single quantum gate is represented by multiple quantum walks, each starting at one of the multiqubit basis states in the qubit domain, see, e.g., Fig.~\ref{fig:2QB} or \ref{fig:2QB-far}.
The problem, therefore, is that of constraint simultaneous optimization of multiple walks, that is, to find parameters of the control field (Rabi frequencies) that define a set of quantum walks satisfying symmetry and time-boundary constraints. When quantum gates span over relatively small number of qubits at a time, this problem is most effectively addressed by investigating analytical solutions (if available), as done in Secs.~\ref{sec:LG}-\ref{sec:SQ}. Specifically, diagonal entangling gates, such as CZ, CCZ, CZZ, and, thus, related CNOT and Tofoli gates, can be obtained by analyzing a set of return quantum walk solutions on simple subgraphs, such as linear-chain graphs of up to 5 states, fan/tree graphs, and square graphs.

Quantum walks solutions to quantum gates are typically not unique and multiple different combinations of control pulse harmonics can accomplish the same gate, as evident from, e.g., solution (\ref{eq:2QB:sol-123}) to a CZ gate. The origin of this is in how a returned walk can occur. A quantum walk on a given graph is guaranteed to return to the starting node if the spectrum of the corresponding adjacency matrix, see, e.g., Eq.~(\ref{eq:LG:CH3:eig}), is composed of integers of the same parity (odd or even). In this case one can chose effective time such that the resulting evolution operator is fully diagonal at the end. As the result, adjacency matrices formed by multicolored control pulses may have different integer spectra and yet produce the same returned walk. Different integer values in the spectra are related to the magnitudes of Rabi frequencies, and, hence, the strength of the control field. This introduces a problem of assessment of gate performance due to different walks: that is, if the control pulse is stronger, it can, in principle, produce faster gates. This speedup however is not usable. In actual physical quantum systems the strength of the control filed and the pulse duration are limited by the hierarchy of transition that may or must not occur to perform desired quantum evolution, as discussed in Sec.~\ref{sec:Systems}. Furthermore, the two always come as a product. Therefore to compare duration of the gates (in real time) due to different solutions, one must rescale control filed amplitudes to keep them within the same limit, see Sec.~\ref{sec:time-comparison}. 
The true, usable gate compression occurs due to more effective utilization of interactions between qubit systems. Specifically, we show that Toffoli gate (CCZ) represented via quantum walks can run almost as fast as a single CNOT (CZ) gate, with the execution time slower only by a factor of $\approx 1.35$. This is in sharp contrast with the standard decomposition theorem,\cite{Markov,nielsenchuang} which states that (three-qubit) Toffoli gates require {\it at least six} CNOT gates.

Not all quantum walk-based solution necessarily lead to speed up of multiqubit gates. Some solutions, particularly, the onces implementing single-qubit gates, can map exactly onto standard $\Lambda$-system single-qubit control commonly (experimentally) performed in many currently available qubit systems. These solutions were outlined in Sec.~\ref{sec:1QB} for completeness. For this reason, it is generally desirable to have analytical solutions to quantum walks on graphs involved in the gate of interest. We have outlined such quantum walk solutions involved in two- and tree-qubit gates in Sec.~\ref{sec:LG}, \ref{sec:TG}, and \ref{sec:SQ}. It is evident, however, that complexity of analytical solutions increases dramatically for larger graphs. This difficulty can be partially avoided by introducing ``classical`` predefined waypoints to guide the walks. This effectively splits the single multicolor pulse into a time sequence of two or several multicolor pulses. The simplification comes at a cost of potentially increasing the overall gate runtime. Yet, we still obtained a significant gate compression for variety of CCZ gates following this procedure, see Sec.~\ref{subsec:sym-B} and Table~\ref{tab:CCZ}.

Finally, we note that this paper is focused on one of the simplest possible scalable quantum registers forming a network of interacting auxiliary states---an infinite chain of qubits. We have also limited our discussion to, at most, three-qubit segments in this register to provide clear examples of walk-based gates. Clearly more developed connectivity should be possible. For example, in quantum dot architecture, discussed in Sec.~\ref{sec:Systems:QDs}, only transitions of $\cal V$-type were used to couple to cavity photons, see Fig.~\ref{fig:QDs}(a) and (b). Additional cavities or cavity modes coupled to these or the other pair of transitions can create nodes accepting more then two connections. The resulting quantum networks will have symmetry of transition amplitudes (adjacency matrix) that is different from the one in the linear chain register. As the result, specific representation of gates via walks obtained in Sec.~\ref{sec:QB} will have to be adjusted to conform to this new symmetry restrictions. Nevertheless, the approach developed in Secs.~\ref{sec:QB}-\ref{sec:SQ} should still be applicable.


\newpage

\begin{appendices}

\section{Exact diagonalization of three-state system}\label{app:LG:CH3}

We include solution obtained from exact diagonalization of a three-states system (chain of three states, Sec.~\ref{sec:LG:CH3}) here for completeness. The adjacency matrix (Hamiltonian) of the system is
\begin{eqnarray}\label{eq:app:LG:CH3:Lambda}
\Lambda =
\left(\begin{array}{ccc}
0 & \Omega_a & 0 \\
\Omega_a^* & 0 & \Omega_b \\
0 & \Omega_b^* & 0
\end{array}\right).
\end{eqnarray}
The evolution operator $U = \exp[-i\tau\Lambda]$, found from the eigenvalue decomposition of $\Lambda$, is
\begin{eqnarray}\label{eq:app:LG:CH3:U}
&&U=
\\\nonumber
&&\left(
{\begin{array}{ccc}
\frac{|\Omega_a|^2+|\Omega_b|^2\cos\tau\sqrt{|\Omega_a|^2+|\Omega_b|^2}}
{|\Omega_a|^2+|\Omega_b|^2}
&
-i\frac{\Omega_b\sin\tau\sqrt{|\Omega_a|^2+|\Omega_b|^2}}
{|\Omega_a|^2+|\Omega_b|^2}
&
-\frac{2\Omega_a\Omega_b\sin^2\frac{\tau\sqrt{|\Omega_a|^2+|\Omega_b|^2}}{2}}
{|\Omega_a|^2+|\Omega_b|^2}
\\
-i\frac{\Omega_b^*\sin\tau\sqrt{|\Omega_a|^2+|\Omega_b|^2}}
{|\Omega_a|^2+|\Omega_b|^2}
&
\cos\tau\sqrt{|\Omega_a|^2+|\Omega_b|^2}
&
-i\frac{\Omega_a\sin\tau\sqrt{|\Omega_a|^2+|\Omega_b|^2}}
{|\Omega_a|^2+|\Omega_b|^2}
\\
-\frac{2\Omega_a^*\Omega_b^*\sin^2\frac{\tau\sqrt{|\Omega_a|^2+|\Omega_b|^2}}{2}}
{|\Omega_a|^2+|\Omega_b|^2}
&
-i\frac{\Omega_a^*\sin\tau\sqrt{|\Omega_a|^2+|\Omega_b|^2}}{|\Omega_a|^2+|\Omega_b|^2} 
& 
\frac{|\Omega_b|^2+|\Omega_a|^2\cos\tau\sqrt{|\Omega_a|^2+|\Omega_b|^2}}
{|\Omega_a|^2+|\Omega_b|^2}
\end{array}}
\right)
\end{eqnarray}
Note that both, the first and the last entry on the diagonal can not be set to $-1$ when $\Omega_a\neq 0$ and $\Omega_b\neq 0$. The central matrix element, however, can become $-1$. The evolution becomes trivial (identity matrix) when $\tau\sqrt{|\Omega_a|^2+|\Omega_b|^2} = 2\pi n$ with $n\in \mathbb{Z}$. When $\tau\sqrt{|\Omega_a|^2+|\Omega_b|^2} = \pi(2n+1)$ evolution started initially from the middle state returns back with a phase of $\pi$.

\section{Eigenvalues of four-state chain adjacency matrix}\label{app:LG:CH4}

The dimensionless adjacency matrix of a chain of four states can be formulated as
\begin{eqnarray}\label{eq:app:LG:CH4:Lambda}
\frac{\tau \Lambda}{\pi} =
\left(\begin{array}{cccc}
0 & a & 0 & 0 \\
a^* & 0 & b & 0 \\
0 & b^* & 0 & c \\
0 & 0 & c^* & 0
\end{array}\right)
\end{eqnarray}
It has four eigenvalues $\pm \lambda_1\tau/\pi$ and $\pm \lambda_2\tau/\pi$, where 
\begin{eqnarray}\label{eq:app:LG:CH4:lambda}
\frac{\lambda_{1,2}\tau}{\pi}
=
\frac{
\sqrt{|a|^2+|b|^2+|c|^2\pm\sqrt{(|a|^2+|b|^2+|c|^2)^2 - 4|a|^2|c|^2}}
}{\sqrt{2}}
\end{eqnarray}
In order to obtain a return walk (evolution) we must set
\begin{eqnarray}\label{eq:app:LG:CH4:condition}
\left\{
\begin{array}{c}
\lambda_1\tau/\pi= n
\\
\lambda_2\tau/\pi=m
\end{array}\right.
, \quad n,m\in {\rm int}
\end{eqnarray}
This produces a system
\begin{eqnarray}\label{eq:app:LG:CH4:solution}
\left\{
\begin{array}{l}
|a|^2+|b|^2+|c|^2=n^2+m^2
\\
\sqrt{(|a|^2+|b|^2+|c|^2)^2 - 4|a|^2|c|^2}
=n^2-m^2
\end{array}\right.
\end{eqnarray}
which can be further simplified to yield Eq.~(\ref{eq:LG:CH4:system}).

\section{Eigenvalues of five-state chain adjacency matrix}\label{app:LG:CH5}

The dimensionless adjacency matrix of a chain of five states is
\begin{eqnarray}\label{eq:LG:CH5:Lambda}
\frac{\Lambda\tau}{\pi} =
\left(\begin{array}{ccccc}
0 & a & 0 & 0 & 0\\
a^* & 0 & b & 0 & 0\\
0 & b^* & 0 & c & 0\\
0 & 0 & c^* & 0 & d\\
0 & 0 & 0 & d^* & 0
\end{array}\right)
\end{eqnarray}
One of the eigenvalues of this matrix is always zero. The other two are $\pm \lambda_1\tau/\pi$ and $\pm \lambda_2\tau/\pi$, where
\begin{eqnarray}\label{eq:app:LG:CH5:ambda}
\lambda_{1,2}\tau/\pi
&=&\frac{
\sqrt{R^2
\pm
\sqrt{
R^4 
- 4|a|^2|c|^2
- 4|b|^2|d|^2
- 4|a|^2|d|^2
}
}
}{\sqrt{2}},
\\\nonumber
R^2 &=&
|a|^2+|b|^2+|c|^2+|d|^2
\end{eqnarray}
Because one of the eigenvalues is always zero, exponentiation of the diagonalized $i\Lambda\tau$ can not produce negative identity matrix. Therefore we must set
\begin{eqnarray}\label{eq:app:LG:CH5:condition}
\left\{
\begin{array}{l}
\lambda_1\tau/\pi= n
\\
\lambda_2\tau/\pi=m
\end{array}\right.
, \quad n,m\in {\rm even}
\end{eqnarray}
This yields the system
\begin{eqnarray}\label{eq:app:LG:CH5:system}
\left\{
\begin{array}{l}
R^2=|a|^2+|b|^2+|c|^2+|d|^2 = n^2 + m^2
\\
\sqrt{R^4 - 4|a|^2|c|^2 - 4|b|^2|d|^2 - 4|a|^2|d|^2} \!=\! n^2\!\! -\! m^2
\end{array}\right.
\end{eqnarray}
which can be further simplified to produce Eq.~(\ref{eq:LG:CH5:system})

\end{appendices}

\end{document}